\documentclass[aps,prb,reprint,amsmath,amssymb]{revtex4-1}
\usepackage[pdftex]{graphicx}
\usepackage{bm}
\begin{document}

\title{Spin-wave thermodynamics of square-lattice antiferromagnets revisited}

\author{Shoji Yamamoto and Yusaku Noriki}

\affiliation{Department of Physics, Hokkaido University,
             Sapporo 060-0810, Japan}

\date{\today}

\begin{abstract}
Modifying the conventional spin-wave theory in a novel manner
based on the Wick decomposition,
we present an elaborate thermodynamics of square-lattice quantum antiferromagnets.
Our scheme is no longer accompanied by the notorious problem of an artificial transition to
the paramagnetic state inherent in modified spin waves in the Hartree-Fock approximation.
In the cases of spin $\frac{1}{2}$ and spin $1$, various modified-spin-wave findings for
the internal energy, specific heat, static uniform susceptibility, and dynamic structure factor
are not only numerically compared with quantum Monte Carlo calculations and Lanczos exact
diagonalizations but also analytically expanded into low-temperature series.
Modified spin waves interacting via the Wick decomposition provide reliable
thermodynamics over the whole temperature range of absolute zero to infinity.
Adding higher-order spin couplings such as ring exchange interaction to the naivest
Heisenberg Hamiltonian, we precisely reproduce inelastic-neutron-scattering measurements of
the high-temperature-superconductor-parent antiferromagnet $\mathrm{La}_2\mathrm{CuO}_4$.
Modifying Dyson-Maleev bosons combined with auxiliary pseudofermions also yields
thermodynamics of square-lattice antiferromagnets free from thermal breakdown, but it is less
precise unless temperature is sufficiently low.
Applying all the schemes to layered antiferromagnets as well, we discuss
the advantages and disadvantages of modified spin-wave and combined boson-pseudofermion
representations.
\end{abstract}

\maketitle
\section{Introduction}\label{S:I}

   Some decades ago, the earliest treatment of antiferromagnetic spin waves (SWs) at finite
temperatures \cite{K568} was modified \cite{T1524,T2494,H4769,T5000} in an attempt to formulate
thermodynamics of square-lattice Heisenberg antiferromagnets and thereby to interpret neutron
scattering measurements on the high-temperature-superconductor-parent compound
$\mathrm{La}_2\mathrm{CuO}_4$
consisting of quantum spins $S=\frac{1}{2}$. \cite{S1613}
Diagonalizing a bosonic Hamiltonian with its sublattice magnetizations constrained to be zero,
Takahashi \cite{T1524,T2494} gave a precise description of the thermal quantities at sufficiently
low temperatures.
Hirsch, Tang, and Lazzouni \cite{H4769,T5000} also demonstrated that such modified SWs (MSWs) well
reproduce exact-diagonalization results for small clusters.
Since we cannot calculate magnetic susceptibilities, whether uniform or staggered and whether
static or dynamic, at finite temperatures within the conventional SW (CSW) theory, their findings
opened up a new avenue for the study of thermodynamics.
However, Takahashi's MSW thermodynamics deteriorates with increasing temperature, not only failing
to design a Schottky-like peak of the specific heat but even encountering an artificial phase
transition of the first order to the trivial paramagnetic solution, \cite{T2494,N034714} similarly
to the Schwinger-boson (SB) mean-field (MF) theory. \cite{A617,Y064426}

   In order to avoid thermal breakdown, Ohara and Yosida \cite{O2521,O3340} proposed another way
of modifying CSWs, which consists of diagonalizing the Hamiltonian without any constraint but
constructing the free energy under vanishing sublattice magnetizations.
This scheme yields a peaked specific heat but spoils the otherwise excellent low-temperature
findings.
Its high-temperature findings are also unfortunate to deviate from the trivial
paramagnetic behavior.
The internal energy and uniform susceptibility per spin never approach zero and
$(g\mu_{\mathrm{B}})^2S(S+1)/3k_{\mathrm{B}}T$, respectively.
Takahashi's MSWs and Auerbach-Arovas' SBs are accompanied by the discontinuous transition but
thereafter stabilized into the correct paramagnetic state, while Ohara-Yosida's MSWs remain
correlated even in the $T\rightarrow\infty$ limit.

   In this context, there is a rather different approach to thermodynamics of layered magnets.
Combining Dyson-Maleev bosons \cite{D1217,D1230} and auxiliary pseudofermions \cite{B351} to
adjust the local Hilbert space dimension to the original spin degrees of freedom, Irkhin, Katanin,
and Katsnelson \cite{I1082} erased the artificial phase transition without spoiling the successful
bosonic description of purely two-dimensional Heisenberg magnets at sufficiently low temperatures.
\cite{T2494,A316,S5028,Y3733}
This formulation unfortunately fails to reproduce the paramagnetic behavior at high temperatures
but gives a satisfactory description of layered systems in the truly critical region crossing their
magnetic ordering temperatures.
It is also noteworthy that a fully convincing description of the two-to-three dimensional crossover
is available within the $1/N$ expansion of the $O(N)$ model, \cite{I12318,I379} i.e. the nonlinear
sigma model generalized to $N$-component spins, rather than through the $1/S$ expansion of any SW
Hamiltonian.

   Under such circumstances, we revisit the SW thermodynamics of
square-lattice antiferromagnets to
find a better solution with particular emphasis on convenience for practical purposes.
Is there something else within a simple spin-wave Hamiltonian that is reliable over the whole
temperature range and applicable to various spins and interactions?
Since the MSW scheme initiated by Takahashi \cite{T1524,T2494} and Hirsch \textit{et al.}
\cite{H4769,T5000} impose a constraint condition of zero staggered magnetization on SWs via
a Bogoliubov transformation dependent on temperature, we refer to this way of modifying CSWs as
a temperature-dependent-diagonalization (TDD)-MSW scheme.
The MSW scheme proposed by Ohara and Yosida \cite{O2521,O3340} manipulates SWs under the same
condition but leaves the CSW Hamiltonian as it is.
Then we refer to this way of modifying CSWs as a temperature-independent-diagonalization (TID)-MSW
scheme.
Modifying sublattice bosons in a TDD manner but bringing them into interaction based on the Wick
decomposition (WD) rather than by the Hartree-Fock (HF) approximation, we can retain the excellent
low-temperature description and connect it naturally with the paramagnetic behavior without any
thermal breakdown.

\section{Modified Spin-Wave Thermodynamics}\label{S:MSWT}

   We divide the square lattice into two sublattices, referred to as A and B, each containing
$N\equiv L/2$ spins of magnitude $S$.
We denote a vector connecting nearest neighbors $\bm{r}_i\ (i\in\mathrm{A})$ and
$\bm{r}_j\ (j\in\mathrm{B})$ by $\bm{\delta}_l$ with $l$ running from $1$ to $z$ to write
the Hamiltonian of our interest as
\begin{equation}
   \mathcal{H}
  =\sum_{i\in\mathrm{A}}\sum_{l=1}^z
   J_{\bm{\delta}_l}\bm{S}_{\bm{r}_i}\cdot\bm{S}_{\bm{r}_i+\bm{\delta}_l}
  =\sum_{j\in\mathrm{B}}\sum_{l=1}^z
   J_{\bm{\delta}_l}\bm{S}_{\bm{r}_j-\bm{\delta}_l}\cdot\bm{S}_{\bm{r}_j}.
   \label{E:H}
\end{equation}
$z$ is equal to $4$ and $J_{\bm{\delta}_l}$ are all set to $J\,(>0)$ unless otherwise noted.
We employ the Dyson-Maleev bosons
\begin{align}
   &
   \left\{
   \!\!
    \begin{array}{l}
     \displaystyle
     S_{\bm{r}_i}^+
    =\sqrt{2S}
     \left(1-\frac{a_{\bm{r}_i}^\dagger a_{\bm{r}_i}}{2S}\right)a_{\bm{r}_i} \\
     S_{\bm{r}_i}^-
    =\sqrt{2S}
     a_{\bm{r}_i}^\dagger \\
     S_{\bm{r}_i}^z
    =S-a_{\bm{r}_i}^\dagger a_{\bm{r}_i} \\
    \end{array}
   \right.\!\!\!,
   \nonumber \\
   &
   \left\{
   \!\!
    \begin{array}{l}
     \displaystyle
     S_{\bm{r}_j}^+
    =\sqrt{2S}
     b_{\bm{r}_j}^\dagger\left(1-\frac{b_{\bm{r}_j}^\dagger b_{\bm{r}_j}}{2S}\right) \\
     S_{\bm{r}_j}^-
    =\sqrt{2S}
     b_{\bm{r}_j} \\
     S_{\bm{r}_j}^z
    =b_{\bm{r}_j}^\dagger b_{\bm{r}_j}-S \\
    \end{array}
   \right.
   \label{E:DMT}
\end{align}
to rewrite the Hamiltonian (\ref{E:H}) into
\begin{align}
   &
   \mathcal{H}=\sum_{l=0}^2\mathcal{H}^{(l)};\ 
   \mathcal{H}^{(2)}
  =-NzJS^2,
   \nonumber \\
   &
   \mathcal{H}^{(1)}
  =JS\sum_{<i,j>}
   (a_{\bm{r}_i}^\dagger a_{\bm{r}_i}+b_{\bm{r}_j}^\dagger b_{\bm{r}_j}
   +a_{\bm{r}_i}^\dagger b_{\bm{r}_j}^\dagger+a_{\bm{r}_i}b_{\bm{r}_j}),\ 
   \nonumber \\
   &
   \mathcal{H}^{(0)}
 =-\frac{J}{2}\sum_{<i,j>}
    a_{\bm{r}_i}^\dagger(a_{\bm{r}_i}+b_{\bm{r}_j}^\dagger)^2 b_{\bm{r}_j}.
   \label{E:HinDMB}
\end{align}
We decompose the $O(S^0)$ quartic Hamiltonian $\mathcal{H}^{(0)}$ into quadratic
(bilinear in the end) terms
\begin{align}
   &\!\!
   \mathcal{H}^{(0)}
  \simeq NzJ\langle\!\langle S-\mathcal{S}\rangle\!\rangle^2
  -J\langle\!\langle S-\mathcal{S}\rangle\!\rangle
   \nonumber \\
   &\!\!\quad\times
   \sum_{<i,j>}
   (a_{\bm{r}_i}^\dagger a_{\bm{r}_i}+b_{\bm{r}_j}^\dagger b_{\bm{r}_j}
   +a_{\bm{r}_i}^\dagger b_{\bm{r}_j}^\dagger+a_{\bm{r}_i}b_{\bm{r}_j})
  \equiv\mathcal{H}^{(0)}_{\mathrm{BL}}
   \label{E:H(0)BL}
\end{align}
to have a tractable SW Hamiltonian,
\begin{align}
   \mathcal{H}
  \simeq\mathcal{H}^{(2)}+\mathcal{H}^{(1)}+\mathcal{H}^{(0)}_{\mathrm{BL}}
  \equiv\mathcal{H}_{\mathrm{BL}},
   \label{E:HBLinDMB}
\end{align}
where we introduce the multivalued double-angle-bracket notation applicable for various
approximation schemes
\begin{align}
   \langle\!\langle\mathcal{S}\rangle\!\rangle
  \equiv
   S
   &
  -\frac{1}{2}
   \langle\!\langle
    a_{\bm{r}_i}^\dagger a_{\bm{r}_i}
   +b_{\bm{r}_i+\bm{\delta}_l}^\dagger b_{\bm{r}_i+\bm{\delta}_l}
   \rangle\!\rangle
   \nonumber \\
   &
  -\frac{1}{2}
   \langle\!\langle
    a_{\bm{r}_i}^\dagger b_{\bm{r}_i+\bm{\delta}_l}^\dagger
   +a_{\bm{r}_i}         b_{\bm{r}_i+\bm{\delta}_l}
   \rangle\!\rangle,
   \label{E:<<S>>}
\end{align}
which we shall read as
the quantum average in the Dyson-Maleev-boson vacuum $\langle\mathcal{S}\rangle_0'$
for the linear SW (LSW) formalism,
the quantum average in the magnon vacuum $\langle\mathcal{S}\rangle_0$
for the WD-based interacting SW (WDISW) formalism,
or
the temperature-$T$ thermal average $\langle\mathcal{S}\rangle_T$
for the HF-decomposition-based interacting SW (HFISW) formalism.
Note that any average $\langle\!\langle\mathcal{S}\rangle\!\rangle$ is independent of the site
indices $\bm{r}_i$ and $\bm{\delta}_l$ by virtue of translation and rotation symmetries.

   Let
$\mathcal{M}_{\mathrm{A}}^\lambda\equiv\sum_{i\in\mathrm{A}}S_{\bm{r}_i}^\lambda$,
$\mathcal{M}_{\mathrm{B}}^\lambda\equiv\sum_{j\in\mathrm{B}}S_{\bm{r}_j}^\lambda$, and
$\mathcal{M}_\pm^\lambda\equiv\mathcal{M}_{\mathrm{A}}^\lambda\pm\mathcal{M}_{\mathrm{B}}^\lambda$.
Then the TDD-MSW theory reads as diagonalizing the effective quadratic Hamiltonian
\begin{align}
   \widetilde{\mathcal{H}}_{\mathrm{BL}}
  \equiv\mathcal{H}_{\mathrm{BL}}+\mu\mathcal{M}_-^z
   \label{E:TDDtildeHBL}
\end{align}
with such $\mu$ as to satisfy $\langle\mathcal{M}_-^z\rangle_T=0$.
We introduce a key variable $p\equiv\sqrt{q^2+1}$ to design various MSWs,
\begin{align}
   &
   p
  \equiv
   \left\{
   \!\!
    \begin{array}{lr}
     \displaystyle
     1-\frac{\mu}{\sum_{l=1}^z J_{\bm{\delta}_l}\langle\!\langle\mathcal{S}\rangle\!\rangle}
    =1-\frac{\mu}{zJ\langle\!\langle\mathcal{S}\rangle\!\rangle} & (\mathrm{TDD}) \\
     \displaystyle
     1                                                           & (\mathrm{TID}) \\
    \end{array}
   \right.\!\!\!.
   \label{E:p}
\end{align}
We further define some functions of $p$,
\begin{align}
   &
   \gamma_{\bm{k}_\nu}
  \equiv
   \frac{\sum_{l=1}^z
         J_{\bm{\delta}_l}\langle\!\langle\mathcal{S}\rangle\!\rangle
         e^{i\bm{k}_\nu\cdot\bm{\delta}_l}}
        {\sum_{l=1}^z
         J_{\bm{\delta}_l}\langle\!\langle\mathcal{S}\rangle\!\rangle}
  =\frac{1}{z}\sum_{l=1}^z e^{i\bm{k}_\nu\cdot\bm{\delta}_l},
   \label{E:gamma}
   \\
   &
   \omega_{\bm{k}_\nu}
  \equiv\frac{\varepsilon_{\bm{k}_\nu}}
             {\sum_{l=1}^z J_{\bm{\delta}_l}\langle\!\langle\mathcal{S}\rangle\!\rangle}
  \equiv\sqrt{p^2-\gamma_{\bm{k}_\nu}^2},
   \label{E:omega}
   \\
   &
   \epsilon
  \equiv\frac{p}{2}-\frac{1}{2N}\sum_{\nu=1}^N\omega_{\bm{k}_\nu},\ 
   \label{E:epsilon}
   \\
   &
   \tau
  \equiv\frac{1}{2N}\sum_{\nu=1}^N\frac{p}{\omega_{\bm{k}_\nu}}-\frac{1}{2}.
   \label{E:tau}
\end{align}
Via the Fourier transformation
\begin{align}
   &
   a_{\bm{k}_\nu}^\dagger
  =\frac{1}{\sqrt{N}}\sum_{i\in\mathrm{A}}
   e^{i\bm{k}_\nu\cdot\bm{r}_i}a_{\bm{r}_i}^\dagger,
   \nonumber \\
   &
   b_{\bm{k}_\nu}
  =\frac{1}{\sqrt{N}}\sum_{j\in\mathrm{B}}
   e^{i\bm{k}_\nu\cdot\bm{r}_j}b_{\bm{r}_j}
   \label{E:FT}
\end{align}
and the Bogoliubov transformation
\begin{align}
   &
   \alpha_{\bm{k}_\nu}^+
  =a_{\bm{k}_\nu}^\dagger\mathrm{sinh}\theta_{\bm{k}_\nu}
  +b_{\bm{k}_\nu}        \mathrm{cosh}\theta_{\bm{k}_\nu},
   \nonumber \\
   &
   \alpha_{\bm{k}_\nu}^-
  =a_{\bm{k}_\nu}        \mathrm{cosh}\theta_{\bm{k}_\nu}
  +b_{\bm{k}_\nu}^\dagger\mathrm{sinh}\theta_{\bm{k}_\nu};
   \nonumber \\
   &
   \mathrm{cosh}2\theta_{\bm{k}_\nu}
  =\frac{p}{\omega_{\bm{k}_\nu}},\ 
   \mathrm{sinh}2\theta_{\bm{k}_\nu}
  =\frac{\gamma_{\bm{k}_\nu}}{\omega_{\bm{k}_\nu}},
   \label{E:BT}
\end{align}
we can diagonalize the effective Hamiltonian into
\begin{align}
   \widetilde{\mathcal{H}}_{\mathrm{BL}}
  =\sum_{l=0}^2 E^{(l)}
  +\sum_{\nu=1}^N
   \varepsilon_{\bm{k}_\nu}
   \sum_{\sigma=\pm}
   \alpha_{\bm{k}_\nu}^{\sigma\dagger}\alpha_{\bm{k}_\nu}^\sigma
  +2\mu NS,
   \label{E:TDDtildeHBLdiag}
\end{align}
where $E^{(2)}$ is the classical ground-state energy and $E^{(l\leq 1)}$ are its
$O(S^l)$ quantum corrections,
\begin{align}
   E^{(2)}
   &
  =-NS^2\sum_{l=1}^z J_{\bm{\delta}_l}
  =-NzJS^2,
   \nonumber
   \\
   E^{(1)}
   &
  =-2NS\epsilon\sum_{l=1}^z J_{\bm{\delta}_l}
  =-2NzJS\epsilon,
   \nonumber \\
   E^{(0)}
   &
  =N\sum_{l=1}^z J_{\bm{\delta}_l}
    \left[
     (S-\langle\!\langle\mathcal{S}\rangle\!\rangle)^2
    +2\epsilon
     (S-\langle\!\langle\mathcal{S}\rangle\!\rangle)
    \right]
   \nonumber
   \\
   &
  =NzJ
   \left[
    (S-\langle\!\langle\mathcal{S}\rangle\!\rangle)^2
   +2(S-\langle\!\langle\mathcal{S}\rangle\!\rangle)\epsilon
   \right].
   \label{E:tildeE(l)}
\end{align}

   Every SW thermodynamics can be formulated in terms of
$\langle\!\langle\mathcal{S}\rangle\!\rangle$ and $p$, which indicate how the SWs are interacting
and modified, respectively.
The TDD-MSW thermal distribution function reads
\begin{align}
   \langle\alpha_{\bm{k}_\nu}^{\sigma\dagger}\alpha_{\bm{k}_\nu}^\sigma\rangle_T
   &
  =\frac
   {\mathrm{Tr}
    \bigl[
     e^{-\varepsilon_{\bm{k}_\nu}
        \alpha_{\bm{k}_\nu}^{\sigma\dagger}\alpha_{\bm{k}_\nu}^\sigma
       /k_{\mathrm{B}}T}
     \alpha_{\bm{k}_\nu}^{\sigma\dagger}\alpha_{\bm{k}_\nu}^\sigma
    \bigr]}
   {\mathrm{Tr}
    \bigl[
     e^{-\varepsilon_{\bm{k}_\nu}
         \alpha_{\bm{k}_\nu}^{\sigma\dagger}\alpha_{\bm{k}_\nu}^\sigma
       /k_{\mathrm{B}}T}
    \bigr]}
   \nonumber \\
   &
  =\frac{1}
   {e^{\varepsilon_{\bm{k}_\nu}/k_{\mathrm{B}}T}-1}
  \equiv
   \bar{n}_{\bm{k}_\nu}
   \label{E:nkTDD}
\end{align}
with $\varepsilon_{\bm{k}_\nu}$ containing $\langle\!\langle\mathcal{S}\rangle\!\rangle$.
Every time we encounter $\langle\!\langle\mathcal{S}\rangle\!\rangle$, we read it according to
the scheme of the time,
\begin{align}
   \langle\mathcal{S}\rangle_0'
   &
  =S
   &
   (\mathrm{LSW}),
   \label{E:<S>'0inDMB}
   \\
   \langle\mathcal{S}\rangle_0
   &
  =S+\epsilon+(p-1)\tau
   &
   (\mathrm{WDISW}),
   \label{E:<S>0inDMB}
   \\
   \langle\mathcal{S}\rangle_T
   &
  =S+\epsilon+(p-1)(\tau+I_2)-I_1
   &
   (\mathrm{HFISW}),
   \label{E:<S>TinDMB}
\end{align}
where the sums
\begin{align}
   &
   I_1
  \equiv\frac{1}{N}\sum_{\nu=1}^N
   \omega_{\bm{k}_\nu}\bar{n}_{\bm{k}_\nu},
   \label{E:I1}
   \\
   &
   I_2
  \equiv\frac{1}{N}\sum_{\nu=1}^N
   \frac{p}{\omega_{\bm{k}_\nu}}\bar{n}_{\bm{k}_\nu}
   \label{E:I2}
\end{align}
still contain $\langle\!\langle\mathcal{S}\rangle\!\rangle$ to be self-consistently determined.
The sublattice magnetizations read
\begin{align}
   &
   \langle\mathcal{M}_{\mathrm{A}}^x\rangle_T
  =\langle\mathcal{M}_{\mathrm{B}}^x\rangle_T
  =0,\ 
   \langle\mathcal{M}_{\mathrm{A}}^y\rangle_T
  =\langle\mathcal{M}_{\mathrm{B}}^y\rangle_T
  =0,
   \nonumber \\
   &
   \langle\mathcal{M}_{\mathrm{A}}^z\rangle_T
 =-\langle\mathcal{M}_{\mathrm{B}}^z\rangle_T
  =N(S-\tau-I_2),
\end{align}
and therefore, the constraint condition is given as
\begin{align}
   \langle\mathcal{M}_-^z\rangle_T
  =2N(S-\tau-I_2)
  =0.
   \label{E:constraintMst}
\end{align}
In modified HFISW (MHFISW) schemes, we solve the simultaneous equations (\ref{E:<S>TinDMB}) plus
(\ref{E:constraintMst}) for $\langle\mathcal{S}\rangle_T$ and $p$.
In modified WDISW (MWDISW) schemes, we substitute Eq. (\ref{E:<S>0inDMB}) into
Eq. (\ref{E:constraintMst}) to obtain $p$ and then $\langle\mathcal{S}\rangle_0$.
In modified LSW (MLSW) schemes, $\langle\mathcal{S}\rangle_0'$ is a constant and therefore we have
only to solve Eq. (\ref{E:constraintMst}) for $p$.
Employing the Bloch-De Dominicis theorem \cite{B459} to evaluate the thermal average of
the quartic Hamiltonian, we can calculate the internal energy,
\begin{align}
   &
   \langle\mathcal{H}^{(2)}\rangle_T
 =-NS^2\sum_{l=1}^z J_{\bm{\delta}_l}
 =-NzJS^2,
   \nonumber
   \\
   &
   \langle\mathcal{H}^{(1)}\rangle_T
 =-2NS\sum_{l=1}^z J_{\bm{\delta}_l}\bigl(\langle\mathcal{S}\rangle_T-S\bigr)
   \nonumber \\
   &\qquad\quad\ 
 =-2NzJS\bigl(\langle\mathcal{S}\rangle_T-S\bigr),
   \nonumber \\
   &
   \langle\mathcal{H}^{(0)}\rangle_T
 =-N\sum_{l=1}^z J_{\bm{\delta}_l}\bigl(\langle\mathcal{S}\rangle_T-S\bigr)^2
   \nonumber \\
   &\qquad\quad\ 
 =-NzJ\bigl(\langle\mathcal{S}\rangle_T-S\bigr)^2;
   \nonumber \\
   &
   E
  \equiv\langle\mathcal{H}\rangle_T
 =-N\sum_{l=1}^z J_{\bm{\delta}_l}\langle\mathcal{S}\rangle_T^2
 =-NzJ\langle\mathcal{S}\rangle_T^2.
   \label{E:<H(l)>T}
\end{align}
Having in mind that
\begin{align}
   e^{ i\widetilde{\mathcal{H}}_{\mathrm{BL}}t/\hbar}
   \alpha_{\bm{k}_\nu}^\sigma
   e^{-i\widetilde{\mathcal{H}}_{\mathrm{BL}}t/\hbar}
  =e^{-i\varepsilon_{\bm{k}_\nu}t/\hbar}
   \alpha_{\bm{k}_\nu}^\sigma,
   \label{E:TimeEvolution}
\end{align}
the dynamic structure factors read
\begin{align}
   &
   S^{\lambda\lambda'}(\bm{q},\omega)
   \nonumber \\
   &\quad
  \equiv\frac{1}{2\pi\hbar L}
   \sum_{k,k'=1}^L
   e^{i\bm{q}\cdot(\bm{r}_k-\bm{r}_{k'})}
   \int_{-\infty}^\infty
   \langle\delta S_{\bm{r}_k}^\lambda(t)\delta S_{\bm{r}_{k'}}^{\lambda'}\rangle_T
   e^{i\omega t}dt;
   \nonumber \\
   &
   \delta S_{\bm{r}_k}^\lambda(t)
  \equiv
   e^{ i\widetilde{\mathcal{H}}_{\mathrm{BL}}t/\hbar}
   \delta S_{\bm{r}_k}^\lambda
   e^{-i\widetilde{\mathcal{H}}_{\mathrm{BL}}t/\hbar},\ 
   \delta S_{\bm{r}_k}^\lambda
  \equiv S_{\bm{r}_k}^\lambda-\langle S_{\bm{r}_k}^\lambda\rangle_T;
   \nonumber \\
   &
   S^{xx}(\bm{q},\omega)
  =S^{yy}(\bm{q},\omega)
  =\frac{S-\tau-I_2}{2}
   (\mathrm{cosh}\theta_{\bm{q}}-\mathrm{sinh}\theta_{\bm{q}})^2
   \nonumber \\
   &\quad\times
   \left[
    \bar{n}_{\bm{q}}\delta(\hbar\omega+\varepsilon_{\bm{q}})
   +(\bar{n}_{\bm{q}}+1)\delta(\hbar\omega-\varepsilon_{\bm{q}})
   \right],
   \nonumber \\
   &
   S^{zz}(\bm{q},\omega)
  =\frac{1}{2N}\sum_{\nu=1}^N
   \biggl\{
    \left[
     \bar{n}_{\bm{k}_\nu}\bar{n}_{\bm{k}_\nu+\bm{q}}
     \delta(\hbar\omega+\varepsilon_{\bm{k}_\nu}+\varepsilon_{\bm{k}_\nu+\bm{q}})
    \right.
   \nonumber \\
   &\quad
    \left.
    +(\bar{n}_{\bm{k}_\nu}+1)(\bar{n}_{\bm{k}_\nu+\bm{q}}+1)
     \delta(\hbar\omega-\varepsilon_{\bm{k}_\nu}-\varepsilon_{\bm{k}_\nu+\bm{q}})
    \right]
   \nonumber \\
   &\quad\times
    \mathrm{sinh}^2(\theta_{\bm{k}_\nu+\bm{q}}-\theta_{\bm{k}_\nu})
   +2\bar{n}_{\bm{k}_\nu}
    (\bar{n}_{\bm{k}_\nu+\bm{q}}+1)
   \nonumber \\
   &\quad\times
    \delta(\hbar\omega+\varepsilon_{\bm{k}_\nu}-\varepsilon_{\bm{k}_\nu+\bm{q}})
    \mathrm{cosh}^2(\theta_{\bm{k}_\nu+\bm{q}}-\theta_{\bm{k}_\nu})
   \biggr\}.
   \label{E:Sqw}
\end{align}
The static structure factors are available from them,
\begin{align}
   S^{\lambda\lambda'}(\bm{q})
   &
  \equiv\frac{1}{L}\sum_{k,k'=1}^L
   e^{i\bm{q}\cdot(\bm{r}_k-\bm{r}_{k'})}
   \langle\delta S_{\bm{r}_k}^\lambda\delta S_{\bm{r}_{k'}}^{\lambda'}\rangle_T
   \nonumber \\
   &
  =\int_{-\infty}^\infty S^{\lambda\lambda'}(\bm{q},\omega)\hbar d\omega;
   \nonumber \allowdisplaybreaks \\
   S^{xx}(\bm{q})
   &
  =S^{yy}(\bm{q})
  =(S-\tau-I_2)
   \nonumber \\
   &\times
   (\mathrm{cosh}\theta_{\bm{q}}-\mathrm{sinh}\theta_{\bm{q}})^2
   \left(\bar{n}_{\bm{q}}+\frac{1}{2}\right),
   \nonumber \allowdisplaybreaks \\
   S^{zz}(\bm{q})
   &
  =\frac{1}{2N}\sum_{\nu=1}^N
   \biggl\{
    \left[
     \bar{n}_{\bm{k}_\nu}\bar{n}_{\bm{k}_\nu+\bm{q}}
    +(\bar{n}_{\bm{k}_\nu}+1)(\bar{n}_{\bm{k}_\nu+\bm{q}}+1)
    \right]
   \nonumber \\
   &\times
    \mathrm{sinh}^2(\theta_{\bm{k}_\nu+\bm{q}}-\theta_{\bm{k}_\nu})
   +2\bar{n}_{\bm{k}_\nu}(\bar{n}_{\bm{k}_\nu+\bm{q}}+1)
   \nonumber \\
   &\times
    \mathrm{cosh}^2(\theta_{\bm{k}_\nu+\bm{q}}-\theta_{\bm{k}_\nu})
   \biggr\},
   \label{E:Sq}
\end{align}
and so are the static uniform susceptibilities,
\begin{align}
   &
   \frac{k_{\mathrm{B}}T}{(g\mu_{\mathrm{B}})^2}
   \chi^{\lambda\lambda}
  \equiv
   \left\langle
   \smash{\left(\mathcal{M}_+^\lambda\right)^2}
   \vphantom{\mathcal{M}_+^\lambda}
   \right\rangle_T
  -\left\langle\mathcal{M}_+^\lambda\right\rangle_T^2
  =LS^{\lambda\lambda}(\bm{0});
   \nonumber \\
   &
   \frac{\chi^{xx}}{L}=\frac{\chi^{yy}}{L}
  =\frac{(g\mu_{\mathrm{B}})^2}{k_{\mathrm{B}}T}
   (S-\tau-I_2)\frac{p-1}{q}\left(\bar{n}_{\bm{0}}+\frac{1}{2}\right),
   \nonumber \\
   &
   \frac{\chi^{zz}}{L}
  =\frac{(g\mu_{\mathrm{B}})^2}{k_{\mathrm{B}}T}I_3;\ 
   I_3
  \equiv
   \frac{1}{N}\sum_{\nu=1}^N \bar{n}_{\bm{k}_\nu}(\bar{n}_{\bm{k}_\nu}+1).
   \label{E:chi}
\end{align}
\begin{figure}
\centering
\includegraphics[width=85mm]{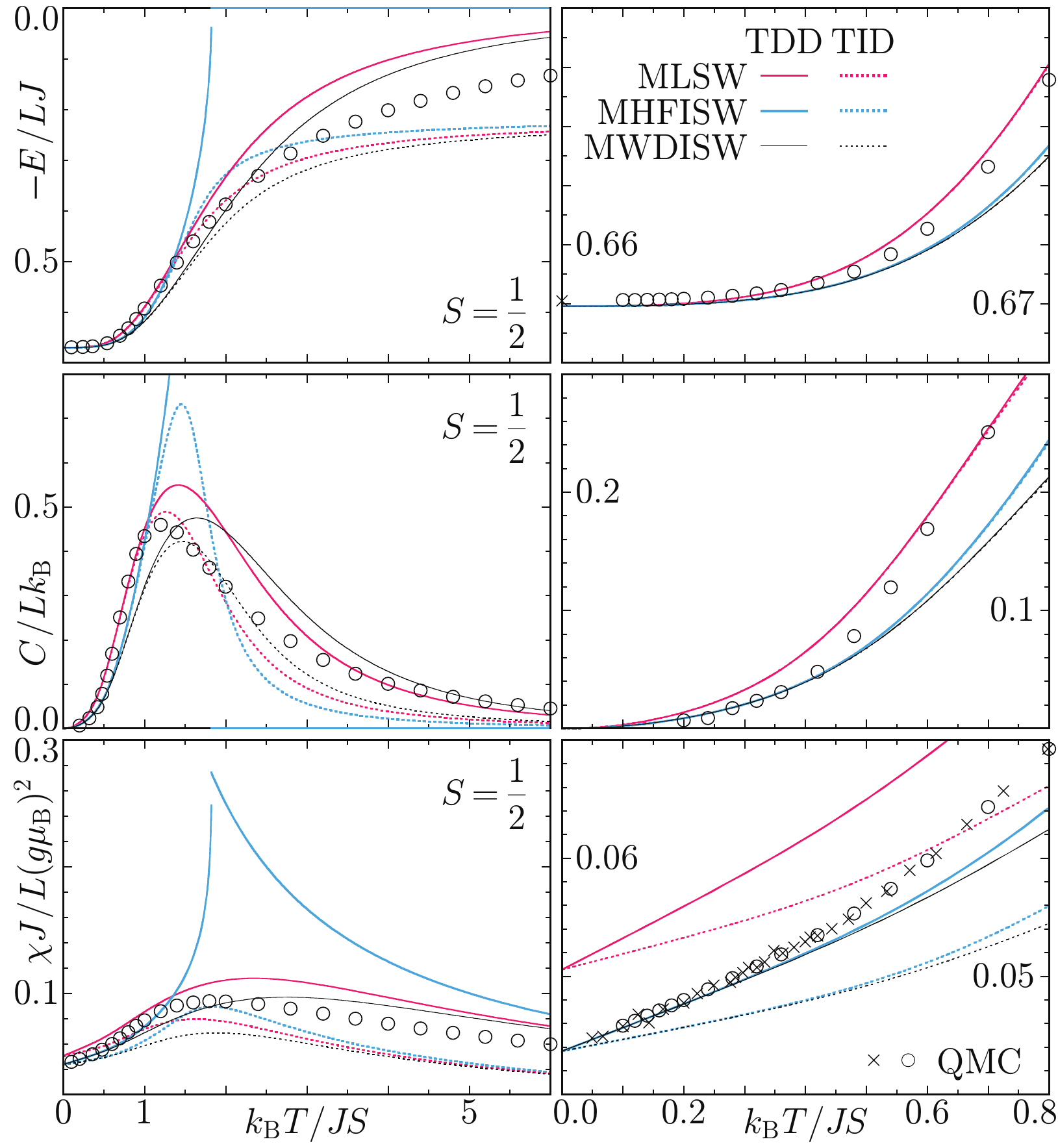}
\vspace*{-2mm}
\caption{(Color online)
         MSW calculations of the internal energy $E$, specific heat $C$ and uniform susceptibility
         $\chi$ as functions of temperature for the Hamiltonian (\ref{E:H}) of $L\rightarrow\infty$
         in comparison with QMC calculations in the case of $S=\frac{1}{2}$.}
\label{F:MSW&QMCofE&C&chi(S=1/2)}
\end{figure}
\begin{figure}
\centering
\includegraphics[width=85mm]{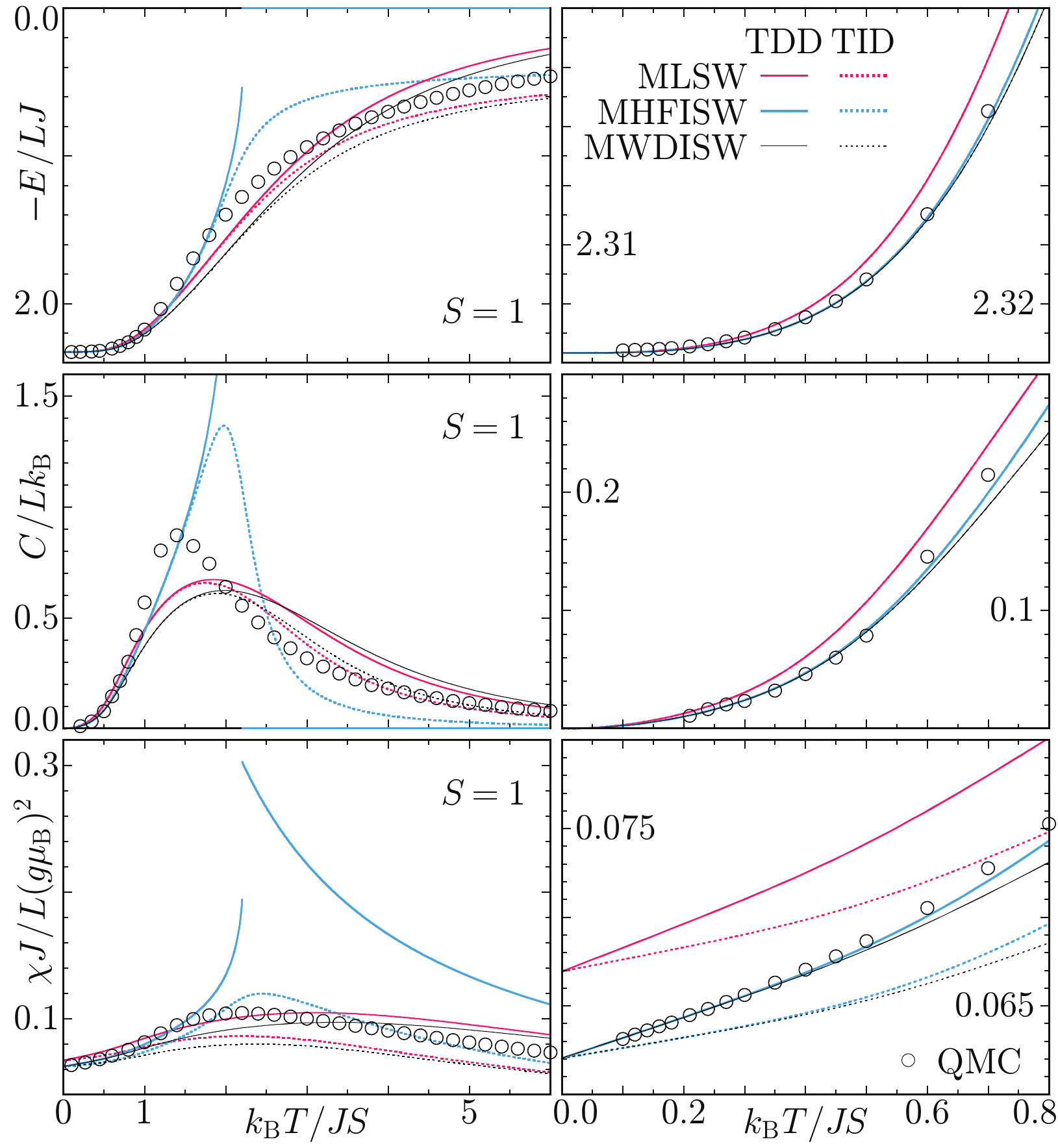}
\vspace*{-2mm}
\caption{(Color online)
         The same as Fig. \ref{F:MSW&QMCofE&C&chi(S=1/2)} in the case of $S=1$.}
\label{F:MSW&QMCofE&C&chi(S=1)}
\end{figure}

   We compare the TDD-MLSW and TDD-MHFISW calculations of
the internal energy $E\equiv\langle\mathcal{H}\rangle_T$,
specific heat $C\equiv\partial E/\partial T$, and
uniform susceptibility $\chi\equiv\sum_{\lambda=x,y,z}\chi^{\lambda\lambda}/3$ with
quantum Monte Carlo (QMC) calculations in Figs. \ref{F:MSW&QMCofE&C&chi(S=1/2)} ($S=\frac{1}{2}$)
and \ref{F:MSW&QMCofE&C&chi(S=1)} ($S=1$), in the former of which Kim and Troyer's QMC findings
\cite{K2705} are also presented ($\times$).
While the TDD-MLSW scheme succeeds in designing antiferromagnetic peaks of $C$ and $\chi$, it is
far from precise at low temperatures, missing the $T\rightarrow 0$ values of $\chi$ considerably.
We can improve these poor low-temperature findings by taking account of the SW interactions
$\mathcal{H}^{(0)}$.
While the thus-obtained TDD-MHFISW findings are highly precise at sufficiently low temperatures,
they completely fail to reproduce the overall temperature dependences.
The worst of them is that they are accompanied by an artificial phase transition of the first
order to the trivial paramagnetic solution at a certain finite temperature.
The specific heat jumps down to zero and the susceptibility switches to that of free spins
$L(g\mu_{\mathrm{B}})^2S(S+1)/3k_{\mathrm{B}}T$ when $k_{\mathrm{B}}T/J$ reaches $0.9108$ and
$2.202$ for $S=\frac{1}{2}$ and $S=1$, respectively, where $p$ diverges, while
$\langle\mathcal{S}\rangle_T$ vanishes, satisfying
$p\langle\mathcal{S}\rangle_T=(k_{\mathrm{B}}T/zJ)\mathrm{ln}(1+1/S)$.

   The TID-MSW theory starts with diagonalizing the CSW Hamiltonian
$\mathcal{H}_{\mathrm{BL}}
\equiv\mathrm{lim}_{\mu\rightarrow 0}\widetilde{\mathcal{H}}_{\mathrm{BL}}$.
\cite{O2521,O3340,Y14008,Y11033}
When we denote the probability of $n$ antiferromagnons of mode $\sigma$ with momentum $\bm{k}_\nu$
emerging at temperature $T$ by $P_{\bm{k}_\nu}(n)$, the free energy reads
\begin{align}
   &
   F
  =\sum_{l=0}^2 E^{(l)}
  +2\sum_{\nu=1}^N
   \varepsilon_{\bm{k}_\nu}
   \sum_{n=0}^\infty nP_{\bm{k}_\nu}(n)
   \nonumber \\
   &\quad
  +2k_{\mathrm{B}}T
   \sum_{\nu=1}^N\sum_{n=0}^\infty
   P_{\bm{k}_\nu}(n)\mathrm{ln}P_{\bm{k}_\nu}(n).
   \label{E:TIDF}
\end{align}
$P_{\bm{k}_\nu}(n)$ is determined so as to minimize the effective free energy
\begin{align}
   &
   \tilde{F}
  =F
  +2\sum_{\nu=1}^N
   \mu_\nu
   \left[
    1-\sum_{n=0}^\infty P_{\bm{k}_\nu}(n)
   \right]
   \nonumber \\
   &\quad
  +2\mu N
   \left[
    S-\tau
   -\frac{1}{N}\sum_{\nu=1}^N\sum_{n=0}^\infty\frac{nP_{\bm{k}_\nu}(n)}{\omega_{\bm{k}_\nu}}
   \right],
   \label{E:TIDtildeF}
\end{align}
where $\mu_\nu$ is obtained from the normalization condition
$\sum_{n=0}^\infty P_{\bm{k}_\nu}(n)=1$, while $\mu$ from the constraint condition
(\ref{E:constraintMst}).
Then we have
\begin{align}
   P_{\bm{k}_\nu}(n)
   &
  =e^{\mu_\nu/k_{\mathrm{B}}T-1}
   e^{-n\tilde{\varepsilon}_{\bm{k}_\nu}/k_{\mathrm{B}}T}
   \nonumber \\
   &
  =\left(
      1-e^{-\tilde{\varepsilon}_{\bm{k}_\nu}/k_{\mathrm{B}}T}
   \right)
   e^{-n\tilde{\varepsilon}_{\bm{k}_\nu}/k_{\mathrm{B}}T}
\end{align}
with the effective energy
\begin{align}
   \tilde{\varepsilon}_{\bm{k}_\nu}
  \equiv\varepsilon_{\bm{k}_\nu}-\frac{\mu}{\omega_{\bm{k}_\nu}}
   \label{E:TIDtildeEk}
\end{align}
to yield the TID-MSW optimal distribution
\begin{align}
   \sum_{n=0}^\infty nP_{\bm{k}_\nu}(n)
   &
  =\frac
   {\mathrm{Tr}
    \bigl[
     e^{-\tilde{\varepsilon}_{\bm{k}_\nu}
         \alpha_{\bm{k}_\nu}^{\sigma\dagger}\alpha_{\bm{k}_\nu}^\sigma/k_{\mathrm{B}}T}
     \alpha_{\bm{k}_\nu}^{\sigma\dagger}\alpha_{\bm{k}_\nu}^\sigma
    \bigr]}
   {\mathrm{Tr}
    \bigl[
     e^{-\tilde{\varepsilon}_{\bm{k}_\nu}
         \alpha_{\bm{k}_\nu}^{\sigma\dagger}\alpha_{\bm{k}_\nu}^\sigma/k_{\mathrm{B}}T}
    \bigr]}
   \nonumber \\
   &
  =\frac{1}
   {e^{\tilde{\varepsilon}_{\bm{k}_\nu}/k_{\mathrm{B}}T}-1}
  \equiv\bar{n}_{\bm{k}_\nu}.
   \label{E:nkTID}
\end{align}
The TID-MSW calculations of $E$, $C$ and $\chi$ are also shown in
Figs. \ref{F:MSW&QMCofE&C&chi(S=1/2)} and \ref{F:MSW&QMCofE&C&chi(S=1)}.
They are indeed free from thermal breakdown but in poor agreement with QMC calculations through
the whole temperature range.
The intermediate-temperature peak of $C$ is too high and the low-temperature increase of $\chi$ is
too slow.
The high-temperature asymptotics is also incorrect
(Refer to Appendix).

   In order to retain the TDD-MHFISW precise low-temperature findings by all means and connect them
naturally with the correct high-temperature asymptotics, we return to the TDD modification scheme
but design interacting SWs in a different manner from the HF approximation.
A new treatment of the $O(S^0)$ quartic Hamiltonian $\mathcal{H}^{(0)}$ consists of applying
the Wick theorem based on the magnon operators $\alpha_{\bm{k}_\nu}^{\sigma\dagger}$ and
$\alpha_{\bm{k}_\nu}^\sigma$ to it and neglecting the residual normal-ordered interaction
$:\!\mathcal{H}^{(0)}\!:$.
Then we have the bilinear Hamiltonian (\ref{E:H(0)BL}) with
$\langle\!\langle\mathcal{S}\rangle\!\rangle$ read as the SW ground-state expectation value
$\langle\mathcal{S}\rangle_0$ given by Eq. (\ref{E:<S>0inDMB}).
We show in Figs. \ref{F:MSW&QMCofE&C&chi(S=1/2)} and \ref{F:MSW&QMCofE&C&chi(S=1)} the MWDISW
calculations as well.
Both TDD and TID MWDISWs describe the Schottky-like peak of $C$ much better than MHFISWs and
their descriptions of $E$ and $C$ are exactly the same at sufficiently low temperatures
[cf. Eq. (\ref{E:E(T)})].
However, a significant difference is detected between the TDD and TID modification schemes
in describing $\chi$.
The low-temperature increase of $\chi$ is imprecisely described by TID MSWs but precisely
reproduced by TDD MWDISWs as well as by TDD MHFISWs [cf. Eq. (\ref{E:chi(T)})].
TDD MHFISWs encounter an artificial phase transition to the paramagnetic solution at an
intermediate temperature, while TDD MWDISWs are free from thermal breakdown.
Unlike TID MSWs, TDD MSWs inherently hit the correct high-temperature limit
(Refer to Appendix).
TDD MHFISWs are unfortunate to artificially jump to the paramagnetic solution, but TDD MLSWs and
TDD MWDISWs are so successful as to give correct high-temperature asymptotics
(See Fig. \ref{F:MSW&MSW+PF&QMCofE&chiT(S=1/2&S=1)logT}).
The TDD-MWDISW thermodynamics is precise at both low and high temperatures and free from any
thermal breakdown.
Let us inquire further into low-temperature MSW findings analytically.

\section{Low-Temperature Series Expansion}\label{S:LTSE}

   In order to convert the $\nu$ summations (\ref{E:I1}), (\ref{E:I2}), and (\ref{E:chi}) into $x$
integrations and thereby expand them into low-temperature series, we define a state-density
function,
\begin{equation}
   w(x)
  =\frac{1}{N}\sum_{\nu=1}^N\delta(x-\omega_{\bm{k}_\nu}+q)\ 
   (0\leq x\leq p-q).
   \label{E:w(x)}
\end{equation}
For the square lattice in the thermodynamic limit, this reads
\begin{align}
   &
   w(x)
  =\left(\frac{2}{\pi}\right)^2
   \frac{(x+q)K\bigl(\sqrt{x^2+2qx}\bigr)}{\sqrt{p^2-(x+q)^2}}
  \equiv 2\sum_{l=0}^\infty w_l x^l;
   \nonumber \\
   &
   K(k)
  \equiv\int_0^{\frac{\pi}{2}}\frac{d\theta}{\sqrt{1-k^2\mathrm{sin}^2\theta}}
  =\frac{\pi}{2}\sum_{n=0}^\infty
   \left[\frac{(2n-1)!!}{(2n)!!}\right]^2
   k^{2n}
   \label{E:w(x)(L->oo)}
\end{align}
with leading coefficients explicitly given as
\begin{align}
   &
   w_0=\frac{q}{\pi},\ 
   w_1=\frac{3q^2+2}{2\pi},\ 
   w_2=\frac{41q^3+36q}{16\pi},
   \nonumber \\
   &
   w_3=\frac{147q^4+164q^2+24}{32\pi},
   \nonumber \\
   &
   w_4=\frac{8649q^5+11760q^3+3280q}{1024\pi},
   \nonumber \\
   &
   w_5=\frac{32307q^6+51894q^4+21168q^2+1312}{2048\pi}.
   \label{E:w_l}
\end{align}
We define integral functions as
\begin{align}
   &
   F_\xi(v,t)
   \nonumber \\
   &
  \equiv
   \left\{
   \!\!
   \begin{array}{lr}
    \displaystyle
    \int_0^{p-q}\!\!\!\!\frac{x^{\xi-1}}{e^{x/t+v}-1}dx
   \mathop{\longrightarrow}^{t\ll p-q}
    \int_0^\infty\frac{t^\xi y^{\xi-1}}{e^{y+v}-1}dy     & (\mathrm{TDD}) \\
    \displaystyle
    \int_0^{p-q}\!\!\!\!\frac{x^{\xi-1}dx}{e^{x/t+tv^2/x}-1}
   \mathop{\longrightarrow}^{t\ll p-q}
    \int_0^\infty\frac{t^\xi y^{\xi-1}dy}{e^{y+v^2/y}-1} & (\mathrm{TID}) \\
   \end{array}
   \right.\!\!\!,
   \label{E:Fxi}
   \\
   &
   G_\xi(v,t)
   \nonumber \\
   &
  \equiv
   \left\{
   \!\!
   \begin{array}{lr}
    \displaystyle
   -\frac{\partial F_\xi(v,t)}{\partial v}
   \mathop{\longrightarrow}^{t\ll p-q}
    \int_0^\infty\frac{t^\xi y^{\xi-1}e^{y+v}}{(e^{y+v}-1)^2}dy
    & (\mathrm{TDD}) \\
    \displaystyle
   -\frac{\partial F_{\xi+1}(v,t)}{\partial(tv^2)}
   \mathop{\longrightarrow}^{t\ll p-q}
    \int_0^\infty\frac{t^\xi y^{\xi-1}e^{y+v^2/y}}{(e^{y+v^2/y}-1)^2}dy
    & (\mathrm{TID}) \\
   \end{array}
   \right.\!\!\!.
   \label{E:Gxi}
\end{align}
If we put
\begin{align}
   &
   f_\xi(v)\equiv\frac{F_\xi(v,t)}{t^\xi\varGamma(\xi)};\ 
   t
  \equiv
   \frac{k_{\mathrm{B}}T}
        {\sum_{l=1}^z J_{\bm{\delta}_l}\langle\!\langle\mathcal{S}\rangle\!\rangle}
  =\frac{k_{\mathrm{B}}T}{zJ\langle\!\langle\mathcal{S}\rangle\!\rangle},\ 
   \nonumber \\
   &
   v
  \equiv
   \left\{
   \!\!
    \begin{array}{lr}
     \displaystyle
     \frac{q}{t}
     & (\mathrm{TDD}) \\
     \displaystyle
     \frac{1}{t}
     \sqrt{\frac{-\mu}{\sum_{l=1}^z J_{\bm{\delta}_l}\langle\!\langle\mathcal{S}\rangle\!\rangle}}
    =\frac{1}{t}
     \sqrt{\frac{-\mu}{zJ\langle\!\langle\mathcal{S}\rangle\!\rangle}}
     & (\mathrm{TID}) \\
    \end{array}
   \right.\!\!\!
   \label{E:t&v}
\end{align}
and assume that $t\ll p-q$, $f_\xi(v)$ for TDD MSWs become Bose-Einstein integral functions and are
therefore expanded in powers of $v$ as
\begin{align}
   f_\xi(v)
  =\frac{(-v)^{\xi-1}}{(\xi-1)!}
   \left(
    \sum_{r=1}^{\xi-1}\frac{1}{r}-\ln{v}
   \right)
  +\sum_{\substack{n=0 \\ n\ne\xi-1}}^{\infty}
   \zeta(\xi-n)
   \frac{(-v)^n}{n!}
\end{align}
for $\xi=1,2,3,\cdots$, while $f_\xi(v)$ for TID MSWs are similarly expanded as
\begin{align}
   &
   \varGamma(\xi)f_\xi(v)
  =(-v^2)^{(\xi-1)/2}
   \left[
    \sum_{r=1}^{(\xi-1)/2}\frac{1}{r}-\ln{v}
   \right]
   \nonumber \\
   &\quad\quad
  +\sum_{\substack{n=0 \\ n\ne(\xi-1)/2}}^{\xi-1}
   \varGamma(\xi-n)
   \zeta(\xi-2n)\frac{(-v^2)^{n}}{n!}
   \nonumber \\
   &\quad\quad
  -\sum_{n=1}^{\infty}\frac{\zeta(2-\xi-2n)}{\varGamma(\xi+n)\varGamma(n)}v^{2\xi+2n-2}
   \left[
    2\varGamma'(1)-2\ln{v}
   \vphantom{\sum_{r=1}^{n-1}}
   \right.
   \nonumber \\
   &\quad\quad\quad
   \left.
   +\frac{\zeta'(2-\xi-2n)}{\zeta(2-\xi-2n)}
   +\sum_{r=1}^{\xi+n-1}\frac{1}{r}
   +\sum_{r=1}^{n-1}\frac{1}{r}
   \right]
\end{align}
for $\xi=1,3,5,\cdots$.

   Now we can expand the integrals (\ref{E:I1}), (\ref{E:I2}), and (\ref{E:chi}) in powers of $t$
and $v$ as
\begin{align}
   &
   I_1
  =2q\sum_{l=0}^\infty\sum_{l'=0}^{1}
   \frac{w_l}{q^{l'}}F_{l+l'+1}(v,t),\ 
   \label{E:I_1int}
   \\
   &
   I_2
  =\left\{
    \!\!
    \begin{array}{lr}
     \displaystyle
     \frac{2p}{q}\sum_{l=0}^\infty\sum_{l'=0}^\infty
     \frac{w_l}{(-q)^{l'}}F_{l+l'+1}(v,t) & (q\neq 0) \\
     \displaystyle
     2\sum_{l=0}^\infty w_l F_l(v,t)      & (q=0)     \\
    \end{array}
   \right.\!\!\!,
   \label{E:I_2int}
   \\
   &
   I_3
  =2\sum_{l=0}^\infty w_l G_{l+1}(v,t).
   \label{E:I_3int}
\end{align}
Having in mind that $p=\sqrt{1+q^2}=\sqrt{1+(vt)^2}$ for TDD MSWs and $p=\sqrt{1+q^2}=1$ for
TID MSWs, we solve Eq. (\ref{E:constraintMst}) for $v$ in an iterative manner to have
\begin{align}
   v=\mathrm{exp}
     \left[
     -\pi\frac{S-\tau_{p=1}}{2t}+\frac{3\zeta(3)}{2}t^2+\frac{123\zeta(5)}{8}t^4+O(t^6)
     \right]
   \label{E:v(t)}
\end{align}
in both cases.
Hence Eqs. (\ref{E:I_1int}) and (\ref{E:I_3int}) read
\begin{align}
   &
   I_1
  =4\frac{\zeta(3)}{\pi}t^3
  +36\frac{\zeta(5)}{\pi}t^5
  +\frac{1845}{2}\frac{\zeta(7)}{\pi}t^7
  +O(t^9),
   \\
   &
   I_3
  =(S-\tau_{p=1})t
  +\left\{
   \!\!
   \begin{array}{lr}
    \displaystyle
    \frac{2t^2}{\pi}\ & (\mathrm{TDD}) \\
    \displaystyle\hphantom{2}
    \frac{ t^2}{\pi}\ & (\mathrm{TID}) \\
   \end{array}
   \right.\!\!\!
   \nonumber \\
   &\quad\ \,
  +6\frac{\zeta(3)}{\pi}t^4
  +123\frac{\zeta(5)}{\pi}t^6
  +O(t^8).
\end{align}
Considering that
$p=1+t^2O(e^{-1/t})$,
$\tau=\tau_{p=1}+tO(e^{-1/t})$,
$\epsilon=\epsilon_{p=1}+t^2O(e^{-1/t})$, and therefore
\begin{align}
   \langle\!\langle\mathcal{S}\rangle\!\rangle
  =\left\{
   \!\!
   \begin{array}{lr}
    S                                                & (\mathrm{LSW})   \\
    S+\epsilon_{p=1}+t^2O(e^{-1/t})                  & (\mathrm{WDISW}) \\
    \displaystyle
    S+\epsilon_{p=1}-4\frac{\zeta(3)}{\pi}t^3+O(t^5) & (\mathrm{HFISW}) \\
   \end{array}
   \right.\!\!\!,
   \label{E:<<S>>(t)}
\end{align}
the nonlinearity of $t$ as a function of $T$ is weak, if any,
\begin{align}
   t
  =\left\{
   \!\!
   \begin{array}{lr}
    \displaystyle
    \frac{k_\mathrm{B}T}{zJS}
    & (\mathrm{LSW}) \\
    \displaystyle
    \frac{k_\mathrm{B}T}{zJ(S+\epsilon_{p=1})}
    \left[1+T^2O(e^{-1/T})\right]
    & (\mathrm{WDISW}) \\
    \displaystyle
    \frac{k_\mathrm{B}T}{zJ(S+\epsilon_{p=1})}
    \left\{
     1
    +\frac{4}{S+\epsilon_{p=1}}\frac{\zeta(3)}{\pi}
     \vphantom{\left[\frac{k_{\mathrm{B}}T}{zJ(S+\epsilon_{p=1})}\right]^3}
    \right. & \\
    \displaystyle\qquad\times
    \left.
     \left[\frac{k_{\mathrm{B}}T}{zJ(S+\epsilon_{p=1})}\right]^3
    +O(T^5)
    \right\}
    & (\mathrm{HFISW}) \\
   \end{array}
   \right.\!\!\!.
   \label{E:t(T)}
\end{align}
We eventually have
\begin{widetext}
\begin{align}
   &\!\!
   \frac{E}{NzJ}
 =-(S+\epsilon_{p=1})^2
  +\left\{
   \!\!
   \begin{array}{lr}
    \displaystyle
    8(S+\epsilon_{p=1})
    \left\{
     \frac{\zeta(3)}{\pi}
     \left(\frac{k_{\mathrm{B}}T}{zJS}\right)^3
    +9\frac{\zeta(5)}{\pi}
     \left(\frac{k_{\mathrm{B}}T}{zJS}\right)^5
    \right\} & \\
    \displaystyle\qquad
   -16\left[\frac{\zeta(3)}{\pi}\right]^2
    \left(\frac{k_{\mathrm{B}}T}{zJS}\right)^6
   +O(T^7)
    & (\mathrm{LSW}) \\
    \displaystyle
    8(S+\epsilon_{p=1})
    \left\{
     \frac{\zeta(3)}{\pi}
     \left[\frac{k_{\mathrm{B}}T}{zJ(S+\epsilon_{p=1})}\right]^3
    +9\frac{\zeta(5)}{\pi}
     \left[\frac{k_{\mathrm{B}}T}{zJ(S+\epsilon_{p=1})}\right]^5
    \right\} & \\
    \displaystyle\qquad
   -16\left[\frac{\zeta(3)}{\pi}\right]^2
    \left[\frac{k_{\mathrm{B}}T}{zJ(S+\epsilon_{p=1})}\right]^6
   +O(T^7)
    & (\mathrm{WDISW}) \\
    \displaystyle
    8(S+\epsilon_{p=1})
    \left\{
     \frac{\zeta(3)}{\pi}
     \left[\frac{k_{\mathrm{B}}T}{zJ(S+\epsilon_{p=1})}\right]^3
    +9\frac{\zeta(5)}{\pi}
     \left[\frac{k_{\mathrm{B}}T}{zJ(S+\epsilon_{p=1})}\right]^5
    \right\} & \\
    \displaystyle\qquad
   +80\left[\frac{\zeta(3)}{\pi}\right]^2
    \left[\frac{k_{\mathrm{B}}T}{zJ(S+\epsilon_{p=1})}\right]^6
   +O(T^7) & (\mathrm{HFISW}) \\
    \end{array}
   \right.\!\!\!,
   \label{E:E(T)}
   \allowdisplaybreaks \\
   &\!\!
   \frac{\chi J}{L(g\mu_{\mathrm{B}})^2}
  =\left\{
   \!\!
   \begin{array}{lr}
    \displaystyle
    \frac{S-\tau_{p=1}}{3zS}
   +\left\{
    \!\!
    \begin{array}{lr}
     \displaystyle
     \frac{2}{3\pi zS}
     \left(\frac{k_{\mathrm{B}}T}{zJS}\right)
     & (\mathrm{TDD}) \\
     \displaystyle
     \frac{1}{3\pi zS}
     \left(\frac{k_{\mathrm{B}}T}{zJS}\right)
     & (\mathrm{TID}) \\
    \end{array}
    \right.\!\!\!
   +\frac{2}{zS}\frac{\zeta(3)}{\pi}
    \left(\frac{k_{\mathrm{B}}T}{zJS}\right)^3
   +O(T^5)
    & (\mathrm{MLSW}) \\
    \displaystyle
    \frac{S-\tau_{p=1}}{3z(S+\epsilon_{p=1})}
   +\left\{
    \!\!
    \begin{array}{lr}
     \displaystyle
     \frac{2}{3\pi z(S+\epsilon_{p=1})}
     \left[\frac{k_{\mathrm{B}}T}{zJ(S+\epsilon_{p=1})}\right]
     & (\mathrm{TDD}) \\
     \displaystyle
     \frac{1}{3\pi z(S+\epsilon_{p=1})}
     \left[\frac{k_{\mathrm{B}}T}{zJ(S+\epsilon_{p=1})}\right]
     & (\mathrm{TID}) \\
    \end{array}
    \right.\!\!\!
    & \\
    \displaystyle\qquad
   +\frac{2}{z(S+\epsilon_{p=1})}\frac{\zeta(3)}{\pi}
    \left[\frac{k_{\mathrm{B}}T}{zJ(S+\epsilon_{p=1})}\right]^3
   +O(T^5)
    & (\mathrm{MWDISW}) \\
    \displaystyle
    \frac{S-\tau_{p=1}}{3z(S+\epsilon_{p=1})}
   +\left\{
    \!\!
    \begin{array}{lr}
     \displaystyle
     \frac{2}{3\pi z(S+\epsilon_{p=1})}
     \left[\frac{k_{\mathrm{B}}T}{zJ(S+\epsilon_{p=1})}\right]
     & (\mathrm{TDD}) \\
     \displaystyle
     \frac{1}{3\pi z(S+\epsilon_{p=1})}
     \left[\frac{k_{\mathrm{B}}T}{zJ(S+\epsilon_{p=1})}\right]
     & (\mathrm{TID}) \\
    \end{array}
    \right.\!\!\!
    & \\
    \displaystyle\qquad
   +\frac{2}{z(S+\epsilon_{p=1})}\frac{\zeta(3)}{\pi}
    \left(1+\frac{2}{3}\frac{S-\tau_{p=1}}{S+\epsilon_{p=1}}\right)
    \left[\frac{k_{\mathrm{B}}T}{zJ(S+\epsilon_{p=1})}\right]^3
   +O(T^4)
    & (\mathrm{MHFISW}) \\
    \end{array}
   \right.\!\!\!.
   \label{E:chi(T)}
\end{align}
\end{widetext}

   There is little difference of $O(e^{-1/T})$ between the TDD-MSW and TID-MSW low-temperature
series expansions of $E$.
Within the first four terms, they are exactly the same not only as each other but also as the CSW
one.
As far as $E$ and therefore $C$ at low temperatures are concerned, it does not matter whether and
how SWs are modified but does whether and how they are interacting.
Note that the chemical potential $v$ has little effect of $O(e^{-1/T})$ on $I_1$.
On the other hand, there is a serious difference of $O(T)$ between the TDD-MSW and TID-MSW
low-temperature series expansions of $\chi$.
While they converge to the same $T\rightarrow 0$ limit, the TID-MSW scheme underestimates
the initial slope by a factor of two.
The TDD-MWDISW and TDD-MHFISW findings are precise and exactly the same as each other within
the first two terms.
With further increasing temperature, the latter deviate from the former and end in the artificial
phase transition to the paramagnetic solution.
We should be reminded that CSWs can reproduce nothing about $\chi$.
$I_3$ expanded in powers of $t$ and $v$ contains a term $\propto -t^2\mathrm{ln}v$, which diverges
in the $v\rightarrow 0$ limit, i.e. within CSW theories, but stays finite by virtue of the constraint
condition (\ref{E:constraintMst}) in MSW theories.

\section{Dynamic Structure Factor}\label{S:DSF}

\begin{figure}
\centering
\includegraphics[width=85mm]{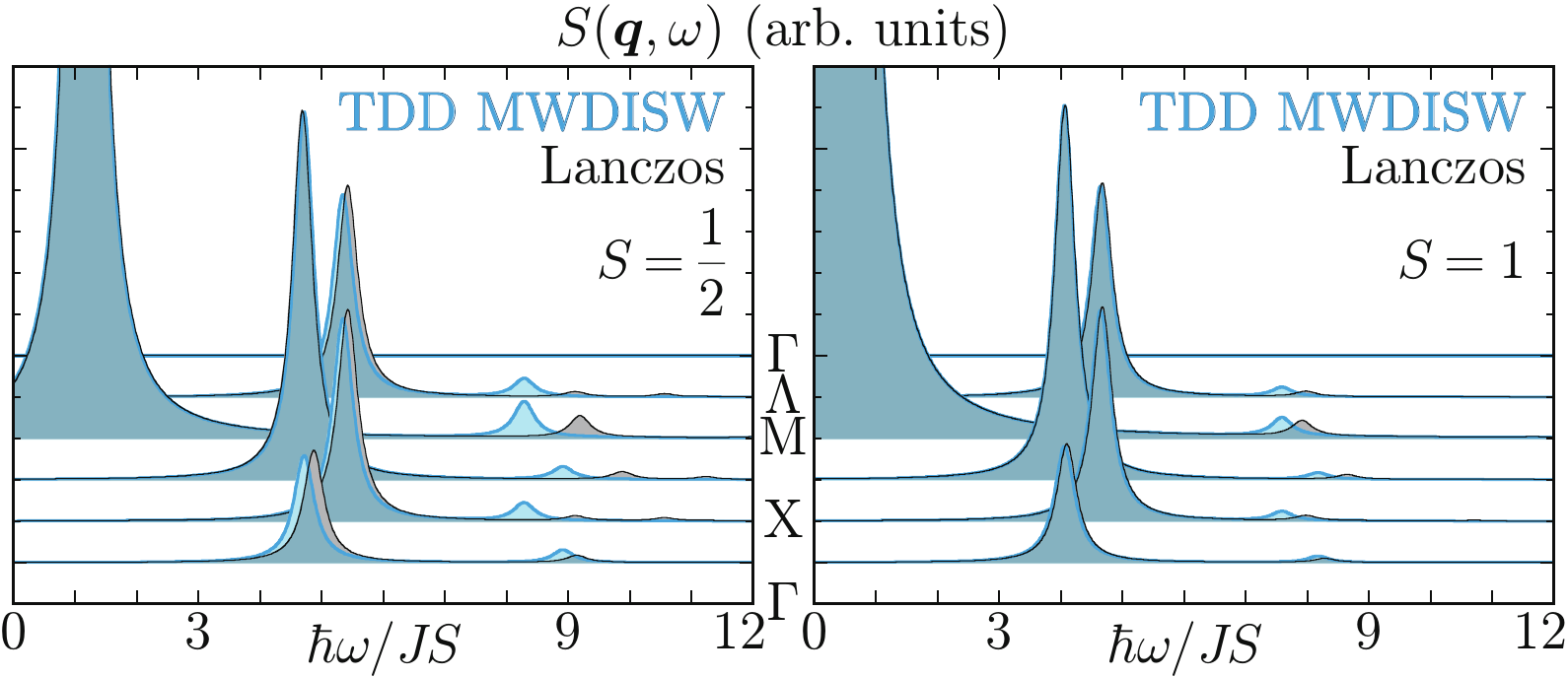}
\vspace*{-2mm}
\caption{(Color online)
         TDD-MWDISW calculations of the dynamic structure factor $S(\bm{q},\omega)$ at absolute
         zero, where MWDISWs and MHFISWs are equivalent,
         for the Hamiltonian (\ref{E:H}) of $L=4^2$ in comparison with Lanczos exact
         diagonalizations in the cases of $S=\frac{1}{2}$ and $S=1$.
         They are originally made of $\delta$-function peaks but Lorentzian-broadened equally for
         comparison.}
\label{F:MSW&LanczosofSqw(S=1/2&S=1)}
\end{figure}

   In order to further demonstrate the quality and reliability of the TDD-MWDISW thermodynamics,
we show in Fig. \ref{F:MSW&LanczosofSqw(S=1/2&S=1)} its findings for the dynamic structure factor
$S(\bm{q},\omega)\equiv\sum_{\lambda=x,y,z}S^{\lambda\lambda}(\bm{q},\omega)/3$ in comparison
with exact calculations of the expression (\ref{E:Sqw}) expanded as a continued fraction.
\cite{G2999,G11766}
TDD MWDISWs give an excellent description especially of the $\delta$-function peaks at
the one-magnon frequency $\varepsilon_{\bm{k}_\nu}$.
There is no visible difference between the MSW and exact calculations of them at all in the case
of $S=1$.
There are two facts noteworthy in this context.
One is that CSWs cannot reproduce anything about $S(\bm{q},\omega)$ even at $T=0$ unless
$L\rightarrow\infty$ and the other is that the TDD-MWDISW and TDD-MHFISW schemes are equivalent
at $T=0$.
In CSW theories, $p=1$ and therefore the excitation energy
$\varepsilon_{\bm{k}_\nu}\equiv zJ\langle\!\langle\mathcal{S}\rangle\!\rangle\omega_{\bm{k}_\nu}$
goes to zero at $\bm{k}_\nu=\bm{0}$.
Then we cannot calculate the $\nu$ summation in Eq. (\ref{E:Sqw}), because
$\mathrm{sinh}2\theta_{\bm{k}_\nu}$ and $\mathrm{cosh}2\theta_{\bm{k}_\nu}$ as well as
$\bar{n}_{\bm{k}_\nu}$ are divergent at $\bm{k}_\nu=\bm{0}$.
If we identify quantum averages at absolute zero, $\langle\ \rangle_{T=0}$, with those in
the magnon vacuum, $\langle\ \rangle_0$, to set $\bar{n}_{\bm{k}}$'s all equal to zero and take
the thermodynamic limit $L\rightarrow\infty$ to convert the intractable $\nu$ summation into
a convergent integral, the \textit{ground-state} structure factors are available within CSW
theories, however the situation is still the same that $\chi\propto S(\bm{0})/T$ is not available
there.
The thus-calculated CSW findings for the ground-state dynamic structure factor are exactly the same
as the $T=0$ MSW calculations at each approximation level, i.e. LSW, WDISW, or HFISW.
Still, the fact remains that we can calculate no magnetic correlation of finite-sized
square-lattice Heisenberg antiferromagnets in terms of SWs without modifying them in a TDD manner.
Since TID MSWs share the same Bogoliubov transformation as CSWs, they are also of little use in
this context.

   $S(\bm{q},\omega)$ of the Hamiltonian (\ref{E:H}) with $S=\frac{1}{2}$ has ever been calculated
in terms of SBs at a MF level \cite{C239} and MHFISWs. \cite{T487}
They yield exactly the same findings at every temperature.
Their findings at $T=0$ are therefore exactly the same as the MWDISW calculations shown in
Fig. \ref{F:MSW&LanczosofSqw(S=1/2&S=1)}.
However, as was fully demonstrated in Figs. \ref{F:MSW&QMCofE&C&chi(S=1/2)} and
\ref{F:MSW&QMCofE&C&chi(S=1)}, MHFISWs are fragile and easy to break down with increasing
temperature, while MWDISWs are robust and free from thermal breakdown.

\section{Comparison with Experiments}\label{S:A}

   It is interesting to analyze inelastic-neutron-scattering (INS) measurements on
the spin-$\frac{1}{2}$ square-lattice antiferromagnet $\mathrm{La}_2\mathrm{CuO}_4$ in terms of our
MWDISW theory.
This material is poorly fitted for the naivest Heisenberg model (\ref{E:H}) but well
describable with a higher-order spin Hamiltonian \cite{C5377} based on a strongly correlated
single-band Hubbard model at half filling, \cite{T1289,M9753,D235130}
\begin{align}
   &
   \mathcal{H}+\mathcal{V}
  =J  (\mathcal{D}_{ \bm{x}        }+\mathcal{D}_{ \bm{y}        })
   \nonumber \\
   &\qquad\quad
  +J' (\mathcal{D}_{ \bm{x}+\bm{y}}+\mathcal{D}_{ \bm{x}-\bm{y}})
  +J''(\mathcal{D}_{2\bm{x}        }+\mathcal{D}_{2\bm{y}        })
   \nonumber \\
   &\qquad\quad
  +J_\square(\mathcal{Q}_{\bm{x},\bm{y},\bm{x}+\bm{y}}
            +\mathcal{Q}_{\bm{y},\bm{x}+\bm{y},\bm{x}}
            -\mathcal{Q}_{\bm{x}+\bm{y},\bm{x},\bm{y}});
   \nonumber \\
   &
   \mathcal{D}_{\bm{\alpha}}
  \equiv\sum_{k=1}^L
   \bm{S}_{\bm{r}_k}\cdot\bm{S}_{\bm{r}_k+\bm{\alpha}},
   \nonumber \\
   &
   \mathcal{Q}_{\bm{\alpha},\bm{\beta},\bm{\gamma}}
  \equiv\sum_{k=1}^L
   (\bm{S}_{\bm{r}_k}\cdot\bm{S}_{\bm{r}_k+\bm{\alpha}})
   (\bm{S}_{\bm{r}_k+\bm{\beta}}\cdot\bm{S}_{\bm{r}_k+\bm{\gamma}}),
   \nonumber \\
   &
   \bm{x}\equiv\bm{\delta}_1=-\bm{\delta}_2,\ 
   \bm{y}\equiv\bm{\delta}_3=-\bm{\delta}_4.
   \label{E:H+V}
\end{align}
Allowing electrons to directly hop only between nearest-neighbor Cu sites and then denoting
the hopping energy and on-site interaction by $t$ and $U$, respectively, we have
\begin{align}
   J=\frac{4t^2}{U}-\frac{24t^4}{U^3},\ 
   J'=J''=\frac{J_\square}{20}=\frac{4t^4}{U^3}.
   \label{E:JJ'J''}
\end{align}
In the expression (\ref{E:H+V}), we regard $\mathcal{H}/U$ as second order in $t/U$ so as to
reproduce the Heisenberg Hamiltonian with a nearest-neighbor-only coupling (\ref{E:H}) and
therefore $\mathcal{V}/U$ as fourth order in $t/U$.
\begin{figure}
\centering
\includegraphics[width=85mm]{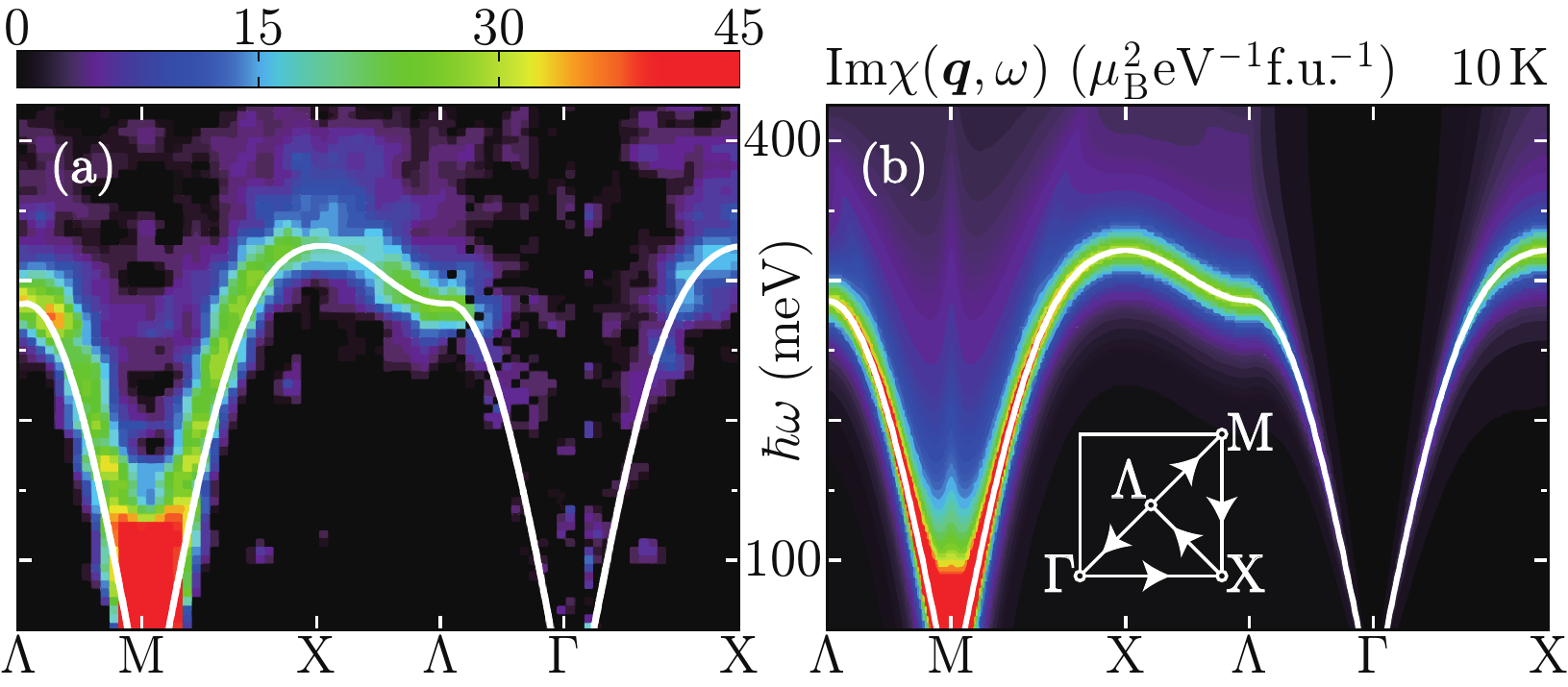}
\vspace*{-2mm}
\caption{(Color online)
         TDD-MWDISW calculations of the dynamic susceptibility $\mathrm{Im}\chi(\bm{q},\omega)$
         for the Hamiltonian (\ref{E:H+V}) of $L\rightarrow\infty$ with $J=145\,\textrm{meV}$
         and $J'=J''=J_\square/20=1.60\,\textrm{meV}$ in the case of $S=\frac{1}{2}$ at
         $T=10\,\mathrm{K}$ (b), whose $\delta$-function peaks are Lorentzian-broadened with
         the use of an incoherent neutron scattering function, \cite{H673} in comparison with
         an INS experiment on $\mathrm{La}_2\mathrm{CuO}_4$ at $T=10\,\mathrm{K}$
         (Ref. \onlinecite{H247001}) (a), where the white guide line is a phenomenological
         dispersion curve obtained by multiplying the CLSW energies with $J=143\pm 2\,\textrm{meV}$
         and $J'=J''=J_\square/20=2.9\pm 0.2\,\textrm{meV}$ by a wavevector-independent quantum
         renormalization factor, $1.18$, deduced from a series-expansion study.
         \cite{S9760}
         The white line in (b) is the up-to-$O(S^0)$ TDD-MWDISW calculation of
         $\varepsilon_{\bm k}$.}
\label{F:MSW&INSofchiqw(S=1/2)T=10K}
\end{figure}
\begin{figure*}
\centering
\includegraphics[width=176mm]{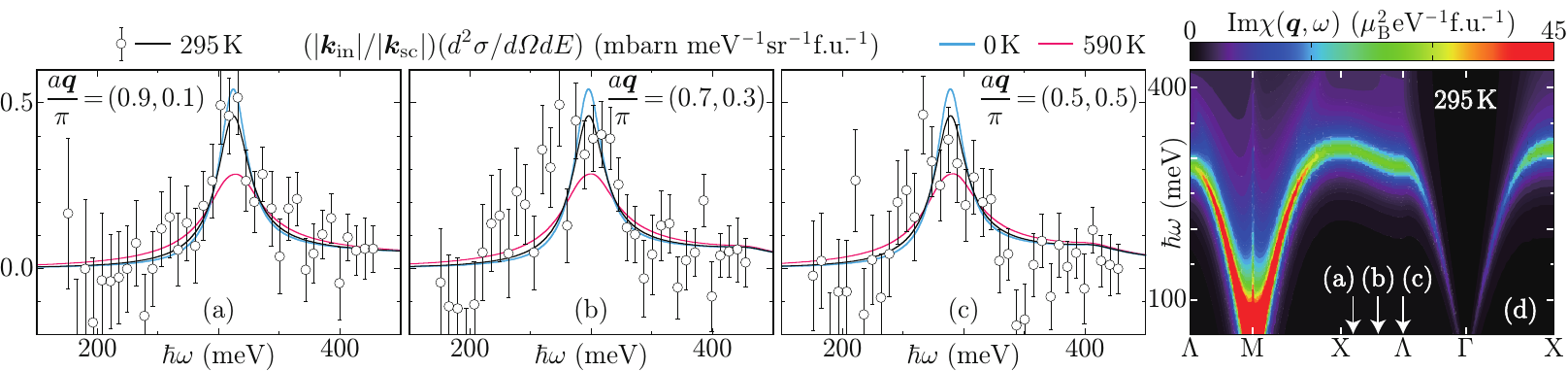}
\vspace*{-2mm}
\caption{(Color online)
         TDD-MWDISW calculations of the double-differential INS cross section
         $d^2\sigma/d\varOmega dE$ (a)--(c) and dynamic susceptibility
         $\mathrm{Im}\chi(\bm{q},\omega)$ (d) for the Hamiltonian (\ref{E:H+V}) of
         $L\rightarrow\infty$ with $J=140\,\textrm{meV}$ and
         $J'=J''=J_\square/20=1.12\,\textrm{meV}$ in the case of $S=\frac{1}{2}$ at various
         temperatures, $T=0,295,590\,\mathrm{K}$, whose $\delta$-function peaks are
         Lorentzian-broadened with the use of an incoherent neutron scattering function.
         \cite{H673}
         Our calculations (a)--(c) are motivated by an INS experiment on
         $\mathrm{La}_2\mathrm{CuO}_4$ at $T=295\,\mathrm{K}$ (Ref. \onlinecite{C5377}) and
         thereby intending to demonstrate their reliability at finite temperatures.}
\label{F:MSW&INSofchiqw(S=1/2)T=295K}
\end{figure*}

   Let $\bm{q}\equiv\bm{k}_{\mathrm{in}}-\bm{k}_{\mathrm{sc}}$ and
$\hbar\omega\equiv E_{\mathrm{in}}-E_{\mathrm{sc}}$ with
$\bm{k}_{\mathrm{in}}$ ($\bm{k}_{\mathrm{sc}}$) and $E_{\mathrm{in}}$ ($E_{\mathrm{sc}}$) being
the initial (final) wavevector and energy of neutrons, respectively.
The double-differential INS cross section, defining the probability of scattering an incident
neutron beam into a particular energy range $dE$ and direction perpendicular to the surface area
subtending the solid angle $d\varOmega$, is directly related to
the imaginary parts of the dynamic susceptibilities $\chi^{\lambda\lambda'}(\bm{q},\omega)$,
\begin{align}
   &
   \frac{d^2\sigma}{d\varOmega dE}
  =L\frac{|\bm{k}_{\mathrm{sc}}|}{|\bm{k}_{\mathrm{in}}|}
   \left(
    \frac{\gamma r_{\mathrm{e}}}{2\mu_{\mathrm{B}}}
   \right)^2
   |F(\bm{q})|^2 e^{-2W(\bm{q})}
   \nonumber \\
   &\quad\times
   \sum_{\lambda=x,y,z}\sum_{\lambda'=x,y,z}
   \left(
    \delta_{\lambda\lambda'}
   -\frac{q_\lambda}{|\bm{q}|}\frac{q_{\lambda'}}{|\bm{q}|}
   \right)
   \frac{\mathrm{Im}\chi^{\lambda\lambda'}(\bm{q},\omega)}
        {1-e^{-\hbar\omega/k_{\mathrm{B}}T}};
   \nonumber \\
   &
   \mathrm{Im}\chi^{\lambda\lambda'}(\bm{q},\omega)
  =(g\mu_{\mathrm{B}})^2
   \left(
    1-e^{-\hbar\omega/k_{\mathrm{B}}T}
   \right)
   S^{\lambda\lambda'}(\bm{q},\omega),
   \label{E:chiqw}
\end{align}
where
$(\gamma r_{\mathrm{e}}/2\mu_{\mathrm{B}})^2$ measures
$72.65\times 10^{-3}\,\mathrm{barn}/\mu_{\mathrm{B}}^2$,
$F(\bm{q})$ is the magnetic form factor defined as
$\int e^{i\bm{q}\cdot\bm{r}}|\phi(\bm{r})|^2 d\bm{r}$ with
$\phi(\bm{r})$ being a Wannier orbital, and
$e^{-2W(\bm{q})}$ is the Debye-Waller factor.
We rewrite  $\mathcal{D}_{\bm{\alpha}}$ and $\mathcal{Q}_{\bm{\alpha},\bm{\beta},\bm{\gamma}}$
in terms of the Dyson-Maleev bosons (\ref{E:DMT}) and denote their $O(S^l)$ terms by
$\mathcal{D}_{\bm{\alpha}}^{(l)}$ and
$\mathcal{Q}_{\bm{\alpha},\bm{\beta},\bm{\gamma}}^{(l)}$, respectively,
similarly to Eq. (\ref{E:HinDMB}).
Decomposing the $O(S^0)$ quartic, $O(S^2)$ quartic, $O(S^1)$ sextic, and $O(S^0)$ octic
interactions $\mathcal{D}_{\bm{\alpha}}^{(0)}$,
$\mathcal{Q}_{\bm{\alpha},\bm{\beta},\bm{\gamma}}^{(2)}$,
$\mathcal{Q}_{\bm{\alpha},\bm{\beta},\bm{\gamma}}^{(1)}$, and
$\mathcal{Q}_{\bm{\alpha},\bm{\beta},\bm{\gamma}}^{(0)}$
all into quadratic terms through the Wick theorem, we evaluate Eq. (\ref{E:chiqw})
in terms of MWDISWs and interpret separate INS experiments on different samples of
$\mathrm{La}_2\mathrm{CuO}_4$ performed
by Headings \textit{et al.} at $T=10\,\mathrm{K}$ \cite{H247001} and
by Coldea \textit{et al.} at $T=295\,\mathrm{K}$. \cite{C5377}
While the Land\'e $g$ factor may depend on direction in a solid, \cite{W1082} here we take an
isotropic $g$ and set it to $2$ for simplicity. \cite{L224511}

   Figure \ref{F:MSW&INSofchiqw(S=1/2)T=10K} shows the experimental and theoretical findings for
the imaginary part of the dynamic susceptibility
$\chi(\bm{q},\omega)\equiv\sum_{\lambda=x,y,z}\chi^{\lambda\lambda}(\bm{q},\omega)$
at $T=10\,\mathrm{K}$.
Supposing the ring exchange interaction is considerably strong, $J_\square=0.41J$, and making
a manual correction to the conventional LSW (CLSW) dispersion relation, Headings \textit{et al.}
\cite{H247001} gave a good guide to the one-magnon cross section
[Fig. \ref{F:MSW&INSofchiqw(S=1/2)T=10K}(a)].
Taking account of the $O(S^0)$ quartic terms $\mathcal{D}_{\bm{\alpha}}^{(0)}$ in the Heisenberg
interactions but discarding the $O(S^1)$ sextic terms
$\mathcal{Q}_{\bm{\alpha},\bm{\beta},\bm{\gamma}}^{(1)}$ and
$O(S^0)$ octic terms $\mathcal{Q}_{\bm{\alpha},\bm{\beta},\bm{\gamma}}^{(0)}$
in the ring exchange interactions, Katanin and Kampf \cite{K100403} demonstrated that conventional
HFISWs (CHFISWs) can indeed reproduce Headings' guiding line with $J=151.9\,\textrm{meV}$ and
$J'=J''=J_\square/9.6=3.80\,\textrm{meV}$.
Their estimate sounds more convincing with a moderate ring exchange interaction,
$J_\square=0.24J$, claiming that any orbital other than Cu $3d_{x^2-y^2}$ should further contribute
to magnetic interactions.
Our full calculation, including all the up-to-$O(S^0)$ terms, can also yield a moderate
ring exchange interaction, $J_\square=20J'=20J''=0.22J$, within the fourth-order perturbation
theory (\ref{E:JJ'J''}).

   Figure \ref{F:MSW&INSofchiqw(S=1/2)T=295K} shows the experimental findings for the INS cross
section $\sigma$ at room temperature in comparison with our theoretical findings for $\sigma$ and
$\chi(\bm{q},\omega)$ at various temperatures.
Coldea \textit{et al.} \cite{C5377} report that the energy dispersion of magnetic excitations along
the high-symmetry directions $\mathrm{X}\,(\pi/a,0)$ to $\Lambda\,(\pi/2a,\pi/2a)$ becomes less
pronounced upon heating.
While Figs. \ref{F:MSW&INSofchiqw(S=1/2)T=10K} and \ref{F:MSW&INSofchiqw(S=1/2)T=295K} look
consistent with such observations, we should be reminded that they are separate observations of
different samples.
The two samples have a small but non-negligible difference of magnetic interaction and that is
mainly why they exhibit a visibly different zone-boundary dispersion.
In this context, we further note that different mechanisms may be responsible for the zone-boundary
dispersion.
Higher-order expansions in $t/U$ and $1/S$ have competing effects on the zone-boundary one-magnon
energies.
Within the LSW description of the naivest Heisenberg Hamiltonian (\ref{E:H}), there is no
dispersion along the zone boundary between $\mathrm{X}$ and $\Lambda$.
Higher-order spin couplings such as ring exchange interaction contribute to raising the one-magnon
energies at around $\mathrm{X}$ with respect to those at around $\Lambda$, \cite{K100403} as was
demonstrated in Figs. \ref{F:MSW&INSofchiqw(S=1/2)T=10K} and \ref{F:MSW&INSofchiqw(S=1/2)T=295K}.
Without any such higher-order exchange coupling, higher-order perturbation corrections within
the Hamiltonian (\ref{E:H}) also make the one-magnon energies along the zone boundary dispersive,
raising those at around $\Lambda$ with respect to those at around $\mathrm{X}$. \cite{S216003,U282}
Such observations are indeed obtained by INS experiments on another spin-$\frac{1}{2}$
square-lattice antiferromagnet, $\mathrm{Cu}(\mathrm{DCOO})_2\cdot 4\mathrm{D}_2\mathrm{O}$,
\cite{R037202,C15264} whose $t/U$ is relatively small with respect to that of
$\mathrm{La}_2\mathrm{CuO}_4$ and hence its nearest-neighbor-only coupling of
$J=6.3\,\mathrm{meV}$. \cite{C561}
There are some other frustrated square-lattice antiferromagnets with dispersive one-magnon energies
along the zone boundary. \cite{K852,T197201}
Dependences of the zone-boundary dispersion on exchange coupling, spin quantum number, and
temperature remain to be investigated from both experimental and theoretical points of view.

   Magnetization measurements on carrier-free $\mathrm{La}_2\mathrm{CuO}_4$
(Ref. \onlinecite{K7430}) and $\mathrm{Cu}(\mathrm{DCOO})_2\cdot 4\mathrm{D}_2\mathrm{O}$
(Refs. \onlinecite{C561},\onlinecite{B883})
reveal their N\'eel transitions at $325\,\mathrm{K}$ and $16.5\,\mathrm{K}$, respectively.
Hence it follows that the experimental observations in Fig. \ref{F:MSW&INSofchiqw(S=1/2)T=295K}
describe the characteristics of SWs in the vicinity of the N\'eel temperature $T_{\mathrm{N}}$.
WDISWs overestimate $T_{\mathrm{N}}$ of layered antiferromagnets, as will be shown in 
Sec. \ref{S:SandD}.
We can make higher-order perturbation corrections to WDISWs in an attempt to reduce their
overestimation of $T_{\mathrm{N}}$ and further interpret various experimental observations of
layered antiferromagnets in the truly critical temperature region near $T_{\mathrm{N}}$.
Such fluctuation corrections, together with auxiliary pseudofermions, indeed improve the HFISW
calculations of sublattice magnetizations in a spin-$1$ layered perovskite with easy-axis
single-ion anisotropy, $\mathrm{K}_2\mathrm{NiF}_4$. \cite{I1082}
In three dimensions, however, the present MSW theories all reduce to a CSW formulation below
$T_{\mathrm{N}}$ with their chemical potential vanishing,
$\mu\rightarrow -0\,(T\rightarrow T_{\mathrm{N}}+0)$.
Therefore, we take more interest in developing an efficient MSW theory in lower dimensions.
The TDD-MWDISW thermodynamics of square-lattice antiferromagnets is precise and analytic at low
temperatures and remains reliable at high temperatures.
We expect that fluctuation corrections will further improve it at intermediate temperatures
rather than attempt to refine that of layered antiferromagnets.

\section{Modified Spin Waves Combined with Pseudofermions}\label{S:MSWwithPF}

   A combined boson-pseudofermion representation of spin operators can also give a reasonable
description of thermodynamic properties.
Tuning this tool in various aspects, Irkhin, Katanin, and Katsnelson \cite{I1082} reproduced
magnetization measurements on a layered antiferromagnet with particular emphasis on
the dimensional-crossover temperature region.
It must be of benefit to our future study to compare MSW theories of current interest with what
they call self-consistent SW (SCSW) theories.
Irkhin-Katanin-Katsnelson's SCSWs within one-particle picture are obtained by combining
TDD MHFISWs with pseudofermions.
Besides them, various MSWs combined with pseudofermions (MSWs+PFs) are available to formulate
thermodynamics.
Before concluding our study, we investigate boson-pseudofermion mixed languages in detail.
Indeed combining TDD MSWs with pseudofermions results in preventing them from thermal breakdown,
but the resultant findings are not necessarily superior to those of pure TDD MSWs.
TDD MLSWs, MHFISWs, and MWDISWs combined with pseudofermions
(MLSWs+PFs, MHFISWs+PFs, and MWDISWs+PFs)
all fail to reproduce the high-temperature paramagnetic behavior correctly.
They remain correlated to underestimate the total spin degrees of freedom even at sufficiently
high temperatures.
Irkhin-Katanin-Katsnelson's SCSWs work better in the vicinity of magnetic ordering than elsewhere
and are therefore suitable for describing two-dimensional magnets with interlayer coupling and/or
magnetic anisotropy, \cite{I1082,G024427} which we shall demonstrate in the final section,
in comparison with purely bosonic TDD-MSW calculations.

   The Bar'yakhtar-Krivoruchko-Yablonski{\u\i} representation of spin operators \cite{B351} reads
\begin{align}
   &
   \left\{
   \!\!
    \begin{array}{l}
     \displaystyle
     S_{\bm{r}_i}^+
    =\sqrt{2S}
     \left[
      1
     -\frac{a_{\bm{r}_i}^\dagger a_{\bm{r}_i}}{2S}
     -\frac{2(2S+1)c_{\bm{r}_i}^\dagger c_{\bm{r}_i}}{2S}
     \right]a_{\bm{r}_i} \\
     S_{\bm{r}_i}^-
    =\sqrt{2S}a_{\bm{r}_i}^\dagger \\
     S_{\bm{r}_i}^z
    =S-a_{\bm{r}_i}^\dagger a_{\bm{r}_i}
    -(2S+1)c_{\bm{r}_i}^\dagger c_{\bm{r}_i} \\
    \end{array}
   \right.\!\!\!,
   \nonumber \\
   &
   \left\{
   \!\!
    \begin{array}{l}
     \displaystyle
     S_{\bm{r}_j}^+
    =\sqrt{2S}b_{\bm{r}_j}^\dagger
     \left[
      1
     -\frac{b_{\bm{r}_j}^\dagger b_{\bm{r}_j}}{2S}
     -\frac{2(2S+1)d_{\bm{r}_j}^\dagger d_{\bm{r}_j}}{2S}
     \right] \\
     S_{\bm{r}_j}^-
    =\sqrt{2S}b_{\bm{r}_j} \\
     S_{\bm{r}_j}^z
    =b_{\bm{r}_j}^\dagger b_{\bm{r}_j}
    +(2S+1)d_{\bm{r}_j}^\dagger d_{\bm{r}_j}-S \\
    \end{array}
   \right.\!\!\!,
   \label{E:BKYT}
\end{align}
where auxiliary pseudofermions, created on sites $\bm{r}_i\ (i\in\mathrm{A})$ and
$\bm{r}_j\ (j\in\mathrm{B})$ by $c_{\bm{r}_i}^\dagger$ and $d_{\bm{r}_j}^\dagger$, respectively,
are employed to bring Dyson-Maleev bosons, created by
$a_{\bm{r}_i}^\dagger$ and $b_{\bm{r}_j}^\dagger$, into kinematic interaction.
The local Hilbert space in which the Bar'yakhtar-Krivoruchko-Yablonski{\u\i} transforms
(\ref{E:BKYT}) operate is spanned by the basis
vectors
\begin{align}
   &
   |n_i,f_i\rangle_{\bm{r}_i}
  \equiv
   \frac{a_{\bm{r}_i}^{\dagger\,n_i}}{\sqrt{n_i!}}
   c_{\bm{r}_i}^{\dagger\,f_i}
   |0\rangle_{\bm{r}_i}\ 
   \nonumber \\
   &\qquad
   (i\in\mathrm{A};\ n_i=0,1,2,\cdots;\ f_i=0,1),
   \nonumber \\
   &
   |n_j,f_j\rangle_{\bm{r}_j}
  \equiv
   \frac{b_{\bm{r}_j}^{\dagger\,n_j}}{\sqrt{n_j!}}
   d_{\bm{r}_j}^{\dagger\,f_j}
   |0\rangle_{\bm{r}_j}\ 
   \nonumber \\
   &\qquad
   (j\in\mathrm{B};\ n_j=0,1,2,\cdots;\ f_j=0,1).
   \label{E:LHSofBKYBwithPF}
\end{align}
Applying $S_{\bm{r}_l}^z$ to each basis set shows that
$|n_l,0\rangle_{\bm{r}_l}$ with $n_l$ larger than $2S$ and $|n_l,1\rangle_{\bm{r}_l}$
are nonphysical states.
The Bar'yakhtar-Krivoruchko-Yablonski{\u\i} bosons combined with pseudofermions (\ref{E:BKYT})
rewrite the Hamiltonian (\ref{E:H}) into
\begin{align}
   &
   \mathcal{H}
  =\sum_{l=0}^2\mathcal{H}^{(l)}
  -J(2S+1)^2\sum_{<i,j>}
   c_{\bm{r}_i}^\dagger c_{\bm{r}_i}d_{\bm{r}_j}^\dagger d_{\bm{r}_j}
   \nonumber \\
   &\quad
  +J(2S+1)\sum_{<i,j>}
   (S-b_{\bm{r}_j}^\dagger b_{\bm{r}_j}-a_{\bm{r}_i}b_{\bm{r}_j})
   c_{\bm{r}_i}^\dagger c_{\bm{r}_i}
   \nonumber \\
   &\quad
  +J(2S+1)\sum_{<i,j>}
   (S-a_{\bm{r}_i}^\dagger a_{\bm{r}_i}-a_{\bm{r}_i}^\dagger b_{\bm{r}_j}^\dagger)
   d_{\bm{r}_j}^\dagger d_{\bm{r}_j}.
   \label{E:HinBKYBwithPF}
\end{align}
Similarly to Eq. (\ref{E:H(0)BL}), we decompose the quartic terms to have a bilinear Hamiltonian,
\begin{align}
   &
   \mathcal{H}
  \simeq
  -NzJS^2
  +NzJ\langle\!\langle S-\mathcal{S}\rangle\!\rangle^2
   \nonumber \\
   &\quad
  +JS\sum_{<i,j>}
   (a_{\bm{r}_i}^\dagger a_{\bm{r}_i}+b_{\bm{r}_j}^\dagger b_{\bm{r}_j}
   +a_{\bm{r}_i}^\dagger b_{\bm{r}_j}^\dagger+a_{\bm{r}_i}b_{\bm{r}_j})
   \nonumber \\
   &\quad
  -J\langle\!\langle S-\mathcal{S}\rangle\!\rangle\sum_{<i,j>}
   (a_{\bm{r}_i}^\dagger a_{\bm{r}_i}+b_{\bm{r}_j}^\dagger b_{\bm{r}_j}
   +a_{\bm{r}_i}^\dagger b_{\bm{r}_j}^\dagger+a_{\bm{r}_i}b_{\bm{r}_j})
   \nonumber \\
   &\quad
  +J\langle\!\langle\mathcal{S}\rangle\!\rangle(2S+1)\sum_{<i,j>}
   (c_{\bm{r}_i}^\dagger c_{\bm{r}_i}+d_{\bm{r}_j}^\dagger d_{\bm{r}_j})
  \equiv\mathcal{H}_{\mathrm{BL}},
   \label{E:HBLinBKYBwithPF}
\end{align}
where the multivalued double-angle-bracket notation is applied to a boson-pseudofermion operator
\begin{align}
   \mathcal{S}
  \equiv
   S
   &
  -\frac{1}{2}
   (a_{\bm{r}_i}^\dagger a_{\bm{r}_i}
   +b_{\bm{r}_i+\bm{\delta}_l}^\dagger b_{\bm{r}_i+\bm{\delta}_l})
   \nonumber \\
   &
  -\frac{1}{2}
   (a_{\bm{r}_i}^\dagger b_{\bm{r}_i+\bm{\delta}_l}^\dagger
   +a_{\bm{r}_i}         b_{\bm{r}_i+\bm{\delta}_l})
   \nonumber \\
   &\!\!\!\!\!\!\!\!\!\!\!\!\!\!
  -\frac{2S+1}{2}
   (c_{\bm{r}_i}^\dagger c_{\bm{r}_i}
   +d_{\bm{r}_i+\bm{\delta}_l}^\dagger d_{\bm{r}_i+\bm{\delta}_l}).
   \label{E:<<S>>inBKYBwithPF}
\end{align}

   In an attempt to formulate thermodynamics in terms of MSWs+PFs, we again introduce the effective
quadratic Hamiltonian
$\widetilde{\mathcal{H}}_{\mathrm{BL}}\equiv\mathcal{H}_{\mathrm{BL}}+\mu\mathcal{M}_-^z$,
similarly to Eq. (\ref{E:TDDtildeHBL}), where $\mu$ is determined so as to satisfy
\begin{align}
   &
   \langle\mathcal{M}_-^z\rangle_T
  \equiv
   \frac
   {\mathrm{Tr}
    \bigl[\mathcal{P}
     e^{-\widetilde{\mathcal{H}}_{\mathrm{BL}}
       /k_{\mathrm{B}}T}
     \mathcal{M}_-^z
    \bigr]}
   {\mathrm{Tr}
    \bigl[\mathcal{P}
     e^{-\widetilde{\mathcal{H}}_{\mathrm{BL}}
       /k_{\mathrm{B}}T}
    \bigr]}
   =0;
   \nonumber \\
   &
   \mathcal{P}
  \equiv
   \mathrm{exp}
   \left[
    i\pi
    \left(
     \sum_{i\in\mathrm{A}}c_{\bm{r}_i}^\dagger c_{\bm{r}_i}
    +\sum_{j\in\mathrm{B}}d_{\bm{r}_j}^\dagger d_{\bm{r}_j}
    \right)
   \right].
   \label{E:constraintMstinBKYBwithPF}
\end{align}
In taking every Bar'yakhtar-Krivoruchko-Yablonski{\u\i} thermal average, the ``pseudoprojection"
operator $\mathcal{P}$ serves to cancel nonphysical-state contributions.
The MSW+PF effective Hamiltonian is diagonalized into
\begin{align}
   &
   \widetilde{\mathcal{H}}_{\mathrm{BL}}
  =\sum_{l=0}^2 E^{(l)}
  +\sum_{\nu=1}^N
   \varepsilon_{\bm{k}_\nu}
   \sum_{\sigma=\pm}
   \alpha_{\bm{k}_\nu}^{\sigma\dagger}\alpha_{\bm{k}_\nu}^\sigma
  +2\mu NS
   \nonumber \\
   &\qquad\ 
  +\delta\varepsilon
   \left(
    \sum_{i\in\mathrm{A}}c_{\bm{r}_i}^\dagger c_{\bm{r}_i}
   +\sum_{j\in\mathrm{B}}d_{\bm{r}_j}^\dagger d_{\bm{r}_j}
   \right),
   \label{E:TDDtildeHBLinBKYBwithPFdiag}
\end{align}
where the MSW energies $\varepsilon_{\bm{k}_\nu}$ look the same as those in
Eq. (\ref{E:TDDtildeHBLdiag}), while pseudofermions, occupying only nonphysical states and
therefore making no physical sense, are immobile to have flat bands above the MSW dispersive bands,
\begin{align}
   \delta\varepsilon
   &
  \equiv
   (2S+1)p
   \sum_{l=1}^z J_{\bm{\delta}_l}\langle\!\langle\mathcal{S}\rangle\!\rangle
   \nonumber \\
   &
  >\sqrt{p^2-\gamma_{\bm{k}_\nu}^2}
   \sum_{l=1}^z J_{\bm{\delta}_l}\langle\!\langle\mathcal{S}\rangle\!\rangle
  \equiv
   \varepsilon_{\bm{k}_\nu}.
   \label{E:deltavarepsilon}
\end{align}
In terms of the Bar'yakhtar-Krivoruchko-Yablonski{\u\i} thermal averages of quasiparticles
\begin{align}
   \bar{n}_{\bm{k}_\nu}
   &
  \equiv
   \langle\alpha_{\bm{k}_\nu}^{\sigma\dagger}\alpha_{\bm{k}_\nu}^\sigma\rangle_T
  =\frac
   {\mathrm{Tr}
    \bigl[\mathcal{P}
     e^{-\widetilde{\mathcal{H}}_{\mathrm{BL}}
       /k_{\mathrm{B}}T}
     \alpha_{\bm{k}_\nu}^{\sigma\dagger}\alpha_{\bm{k}_\nu}^\sigma
    \bigr]}
   {\mathrm{Tr}
    \bigl[\mathcal{P}
     e^{-\widetilde{\mathcal{H}}_{\mathrm{BL}}
       /k_{\mathrm{B}}T}
    \bigr]}
   \nonumber \\
   &
  =\frac{1}
   {e^{\varepsilon_{\bm{k}_\nu}/k_{\mathrm{B}}T}-1},
   \label{E:nkTDDinBKYBwithPF}
   \\
   \bar{f}
   &
  \equiv
   \langle f_{\bm{r}_k}^\dagger f_{\bm{r}_k}\rangle_T
  =\frac
   {\mathrm{Tr}
    \bigl[\mathcal{P}
     e^{-\widetilde{\mathcal{H}}_{\mathrm{BL}}
       /k_{\mathrm{B}}T}
     f_{\bm{r}_k}^\dagger f_{\bm{r}_k}
    \bigr]}
   {\mathrm{Tr}
    \bigl[\mathcal{P}
     e^{-\widetilde{\mathcal{H}}_{\mathrm{BL}}
       /k_{\mathrm{B}}T}
    \bigr]}
   \nonumber \\
   &
  =\frac{1}
   {e^{\delta\varepsilon/k_{\mathrm{B}}T-i\pi}+1};\ 
   f_{\bm{r}_k}
  \equiv
   \left\{
   \!\!
    \begin{array}{lr}
     \displaystyle
     c_{\bm{r}_k} & (k\in\mathrm{A}) \\
     \displaystyle
     d_{\bm{r}_k} & (k\in\mathrm{B}) \\
    \end{array}
   \right.\!\!\!,
   \label{E:fTDDinBKYBwithPF}
\end{align}
the order parameter $\langle\!\langle\mathcal{S}\rangle\!\rangle$ is explicitly written as
\begin{align}
   \langle\mathcal{S}\rangle_0'
   &
  =S
   \hspace{-7.25em}
   &
   (\mbox{LSW+PF}),
   \label{E:<S>'0inBKYBwithPF}
   \\
   \langle\mathcal{S}\rangle_0
   &
  =S+\epsilon+(p-1)\tau
   \hspace{-7.25em}
   &
   (\mbox{WDISW+PF}),
   \label{E:<S>0inBKYBwithPF}
   \\
   \langle\mathcal{S}\rangle_T
   &
  =S+\epsilon+(p-1)(\tau+I_2)-I_1-(2S+1)\bar{f}
   \hspace{-7.25em}
   &
   \nonumber \\
   &
   \hspace{-7.25em}
   &
   (\mbox{HFISW+PF}).
   \label{E:<S>TinBKYBwithPF}
\end{align}
The sublattice magnetizations read
\begin{align}
   &
   \langle\mathcal{M}_{\mathrm{A}}^x\rangle_T
  =\langle\mathcal{M}_{\mathrm{B}}^x\rangle_T
  =0,\ 
   \langle\mathcal{M}_{\mathrm{A}}^y\rangle_T
  =\langle\mathcal{M}_{\mathrm{B}}^y\rangle_T
  =0,
   \nonumber \\
   &
   \langle\mathcal{M}_{\mathrm{A}}^z\rangle_T
 =-\langle\mathcal{M}_{\mathrm{B}}^z\rangle_T
  =N[S-\tau-I_2-(2S+1)\bar{f}],
\end{align}
and therefore, the constraint condition is given as
\begin{align}
   \langle\mathcal{M}_-^z\rangle_T
  =2N[S-\tau-I_2-(2S+1)\bar{f}]
  =0.
   \label{E:constraintMstinBKYBwithPF}
\end{align}
We choose one from Eqs. (\ref{E:<S>'0inBKYBwithPF})--(\ref{E:<S>TinBKYBwithPF}) and solve it
simultaneously with Eq. (\ref{E:constraintMstinBKYBwithPF}) for
$\langle\!\langle\mathcal{S}\rangle\!\rangle$ and/or $p$, where every time we encounter
$\langle\!\langle\mathcal{S}\rangle\!\rangle$, we read it as one of
Eqs. (\ref{E:<S>'0inBKYBwithPF})--(\ref{E:<S>TinBKYBwithPF}) according to the scheme of the time.
The internal energy and uniform susceptibilities are expressed as
\begin{align}
   &
   E
 =-N\sum_{l=1}^z J_{\bm{\delta}_l}\langle\mathcal{S}\rangle_T^2
 =-NzJ\langle\mathcal{S}\rangle_T^2,
   \label{E:EinBKYBwithPF}
   \\
   &
   \frac{\chi^{xx}}{L}=\frac{\chi^{yy}}{L}
  =\frac{(g\mu_{\mathrm{B}})^2}{k_{\mathrm{B}}T}
   [S-\tau-I_2-(2S+1)\bar{f}]
   \nonumber \\
   &\qquad\times
   \frac{p-1}{q}\left(\bar{n}_{\bm{0}}+\frac{1}{2}\right),
   \nonumber \\
   &
   \frac{\chi^{zz}}{L}
  =\frac{(g\mu_{\mathrm{B}})^2}{k_{\mathrm{B}}T}[I_3+(2S+1)^2\bar{f}(1-\bar{f})].
   \label{E:chiinBKYBwithPF}
\end{align}

\begin{figure}
\centering
\includegraphics[width=85mm]{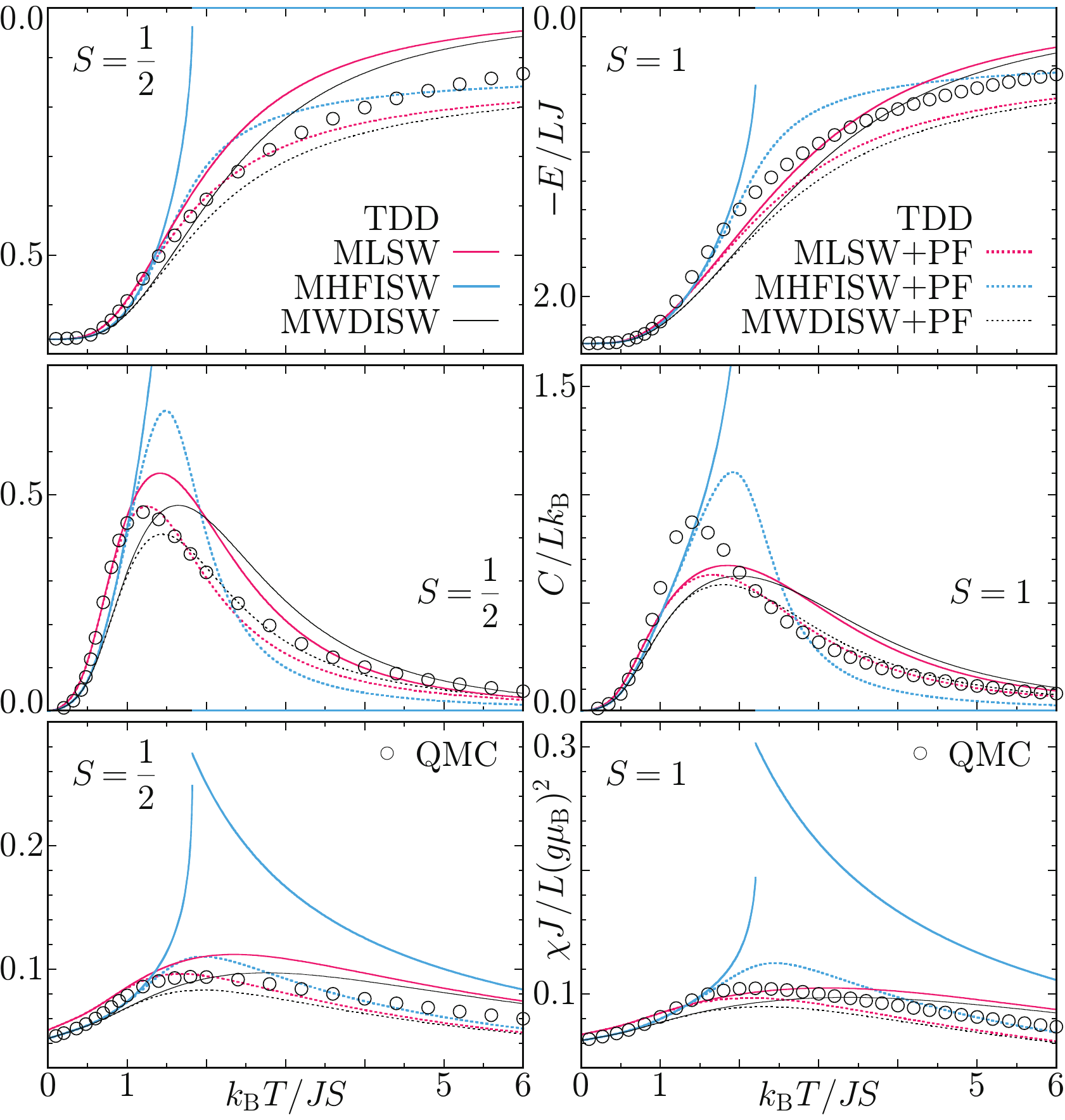}
\vspace*{-2mm}
\caption{(Color online)
         TDD-MSW+PF calculations of the internal energy $E$, specific heat $C$ and uniform
         susceptibility $\chi$ as functions of temperature for the Hamiltonian (\ref{E:H}) of
         $L\rightarrow\infty$ in comparison with TDD-MSW and QMC calculations in the cases of
         $S=\frac{1}{2}$ and $S=1$.}
\label{F:MSW&MSW+PF&QMCofE&C&chiT(S=1/2&S=1)}
\end{figure}

   We compare the thus-calculated $E$, $C$, and $\chi$ of the purely two-dimensional
square-lattice Heisenberg antiferromagnets (\ref{E:H}) with the TDD-MSW calculations in
Fig. \ref{F:MSW&MSW+PF&QMCofE&C&chiT(S=1/2&S=1)}.
MHFISWs and MWDISWs, whether combined with pseudofermions or not, bring exactly the same results at
sufficiently low temperatures, namely, within the first three (up to $T^5$) terms of $E$ and
the first two (up to $T^2$) terms of $\chi T$.
With further increasing temperature, there occurs a difference between their calculations.
The MHFISW findings increase too rapidly as functions of temperature and encounter an
artificial transition to the paramagnetic solution without auxiliary pseudofermions.
Their first-order transition is fictitious indeed, but they hit the correct high-temperature limit.
From this point of view, MHFISWs+PFs unfortunately mishit the high-temperature paramagnetic
behavior (Refer to Appendix), even though they are free from thermal
breakdown.
In the TDD-MHFISW+PF formulation, every thermodynamic quantity is overestimated and underestimated
at intermediate and high temperatures, respectively.
The internal energy and uniform susceptibility per spin never approach zero and
$(g\mu_{\mathrm{B}})^2S(S+1)/3k_{\mathrm{B}}T$, respectively, even when temperature rises.
This behavior is reminiscent of the TID-MSW findings.

   Now we are in a position to have a bird's-eye view of various SW theories
designed for thermodynamics of square-lattice antiferromagnets.
Whether they are useful will depend on how we use them.
If we take interest only in the low-temperature properties, TDD MSWs, no matter how they are
interacting and whether or not they are combined with PFs, may serve our purpose.
If we are under the necessity of reproducing the overall temperature dependences of experimental
findings, TDD MHFISWs are no longer useful at all but TDD MWDISWs remain reliable.
Indeed TDD MHFISWs are inferior in describing the thermodynamic properties of low-dimensional
antiferromagnets, but they are potentially superior in predicting the ordering temperature of
three-dimensional layered antiferromagnets
(cf. Fig. \ref{F:MSW&MSW+PFofE&Mst(S=1/2&S=1)3D} and see Ref. \onlinecite{I1082} as well).
While TID MSWs appear to have no particular merit, there is an example of their successful
detection \cite{Y074703} of novel nuclear spin relaxation mediated by multimagnons
\cite{P398,B359} in an intertwining double-chain ferrimagnet.
It is hard to say in a word which scheme is the best.
When we investigate two-dimensional antiferromagnets with the aim to quantitatively analyze
their thermodynamic properties as functions of temperature,
Table \ref{T:YesNoEvaluation} serves to evaluate various MSW representations.
\begin{table}
\caption{Evaluation of thermodynamics of square-lattice Heisenberg
         antiferromagnets obtained through various formulations from the points of view of
         whether it is  precise, valid, and correct at low, intermediate, and high temperatures,
         respectively.
         We judge the high-temperature asymptotics of the TDD-MHFISW thermodynamics to be
         incorrect in spite of its correct high-temperature limit, because TDD MHFISWs
         artificially jump to the paramagnetic phase at a finite temperature.}
\begin{tabular}{lccc}
\hline \hline
\\[-3.5mm]
              & \parbox{16.75mm}{Precise \\ low-$T$ \\ analytics     }
              & \parbox{16.75mm}{Free from \\ thermal \\ breakdown   }
              & \parbox{16.75mm}{Correct \\ high-$T$ \\ asymptotics  } \\
\\[-3.5mm]
\hline
TID MLSW      & {\scriptsize No} & Yes              & {\scriptsize No} \\
TID MWDISW    & {\scriptsize No} & Yes              & {\scriptsize No} \\
TID MHFISW    & {\scriptsize No} & Yes              & {\scriptsize No} \\
TDD MLSW      & {\scriptsize No} & Yes              & Yes              \\
TDD MWDISW    & Yes              & Yes              & Yes              \\
TDD MHFISW    & Yes              & {\scriptsize No} & {\scriptsize No} \\
TDD MLSW+PF   & {\scriptsize No} & Yes              & {\scriptsize No} \\
TDD MWDISW+PF & Yes              & Yes              & {\scriptsize No} \\
TDD MHFISW+PF & Yes              & Yes              & {\scriptsize No} \\
\hline \hline
\end{tabular}
\label{T:YesNoEvaluation}
\end{table}

\section{Summary and Discussion}\label{S:SandD}

   TDD MWDISWs serve as a sophisticated language for extensive quantum antiferromagnets in two
dimensions.
Their thermodynamics is highly precise at low temperatures, sufficiently robust against
thermalization, and well applicable to frustrated antiferromagnets as well, covering both static
and dynamic properties.
The spin-$\frac{1}{2}$ square-lattice Heisenberg antiferromagnet (\ref{E:H}) frustrated by
diagonal nearest-neighbor couplings have also ever been studied in terms of MSWs by numerous
authors with particular emphasis on its nature at absolute zero.
Chandra and Doucot \cite{C9335} discussed within the CLSW scheme that a spin liquid phase may exist
in the sense of the conventional antiferromagnetic long-range order disappearing.
Hirsch and Tang \cite{H2887} employed the TDD-MLSW scheme in an attempt to reveal
a transition to a disordered phase.
Nishimori and Saika \cite{N4454} followed to introduce the TDD-MHFISW scheme and claim that
a N\'eel-type long-range order consisting of either two or four sublattices should survive
the strong frustration.
While there are analytic \cite{I8206} and numerical \cite{Y2384,W1350021} MSW calculations of
static properties at finite temperatures as well, the TDD-MWDISW thermodynamics of this model is
never yet revealed.
Such calculation is not only intriguing in itself but may also give a piece of information
otherwise unavailable.

   We take a particular interest in developing a SW thermodynamics unattainable within
the CSW theory.
We therefore devote much effort to describing low-dimensional magnets with no long-range order
at finite temperatures.
In three dimensions, a long-range order survives thermalization more or less, during which every
MSW scheme reduces to a CSW formulation with its chemical potential unavailable from
the constraint condition (\ref{E:constraintMst}). \cite{R119}
In lower dimensions, SWs may be modified over the whole temperature range to have their otherwise
unavailable thermodynamics.
However, real layered magnets encounter an order-disorder phase transition at a finite temperature,
which we shall denote by $T_{\mathrm{N}}$, due to weak interlayer coupling and/or magnetic
anisotropy.
In an attempt to interpret the spontaneous magnetization as an anomalous function of temperature
in a three-dimensional ferrimagnet, Karchev \cite{K325219,K216003} proposed modifying CSWs
with a temperature- and sublattice-dependent constraint condition.
Focusing on a narrow critical region near $T_{\mathrm{N}}$, Irkhin, Katanin, and Katsnelson
\cite{I1082} applied a MSW theory refined with auxiliary pseudofermions to quasi-two-dimensional
ferromagnets and antiferromagnets.
There is also a possibility of further developing our scheme along this line.

   In order to gain better understanding of SW thermodynamics, we finally compare the TDD-MSW and
TDD-MSW+PF findings for the order-disorder phase transitions of layered Heisenberg antiferromagnets
describable within the Hamiltonian (\ref{E:H}) of $z=6$.
When we set $J_{\bm{\delta}_l}$ to $J\,(>0)$ and $J_\perp\,(>0)$ for $1\leq l\leq 4$ and
$5\leq l\leq 6$, respectively, the Hamiltonian of current interest reads
\begin{align}
   \mathcal{H}
  =J(\mathcal{D}_{\bm{x}}+\mathcal{D}_{\bm{y}})+J_\perp\mathcal{D}_{\bm{z}};\ 
   \bm{z}\equiv\bm{\delta}_5=-\bm{\delta}_6.
   \label{E:H+Hperp}
\end{align}
In terms of the Dyson-Maleev bosons (\ref{E:DMT}) and Bar'yakhtar-Krivoruchko-Yablonski{\u\i}
bosons combined with pseudofermions (\ref{E:BKYT}), this Hamiltonian is also rewritten into
(\ref{E:HinDMB}) and (\ref{E:HinBKYBwithPF}) and then decomposed into the quadratic forms
(\ref{E:HBLinDMB}) and (\ref{E:HBLinBKYBwithPF}), respectively.
Under the constraint condition of zero staggered magnetization, the MSW and MSW+PF effective
quadratic Hamiltonians
$\widetilde{\mathcal{H}}_{\mathrm{BL}}\equiv\mathcal{H}_{\mathrm{BL}}+\mu\mathcal{M}_-^z$
are diagonalized into (\ref{E:TDDtildeHBLdiag}) and (\ref{E:TDDtildeHBLinBKYBwithPFdiag}),
respectively.
The bond-dependent order parameters $\langle\!\langle\mathcal{S}\rangle\!\rangle$ are
explicitly written as
\begin{align}
   \langle\mathcal{S}\rangle_0'
   &
  =S
   \hspace{-41mm}
   &
   (\mbox{LSW},\,\mbox{LSW+PF}),
   \label{E:<Sdelta_l>'0inDMB&BKYBwithPF}
   \\
   \langle\mathcal{S}\rangle_0
   &
  =S+\epsilon+(p-1)\tau
   \hspace{-41mm}
   &
   \nonumber \\
   &
  +\eta'-\eta
   \hspace{-41mm}
   &
   (\mbox{WDISW},\,\mbox{WDISW+PF}),
   \label{E:<Sdelta_l>0inDMB&BKYBwithPF}
   \\
   \langle\mathcal{S}\rangle_T
   &
  =S+\epsilon+(p-1)(\tau+I_2)-I_1
   \hspace{-41mm}
   &
   \nonumber \\
   &
  +\eta'-\eta+H'-H
   \hspace{-41mm}
   &
   (\mbox{HFISW}),
   \nonumber \\
   \langle\mathcal{S}\rangle_T
   &
  =S+\epsilon+(p-1)(\tau+I_2)-I_1-(2S+1)\bar{f}
   \hspace{-41mm}
   &
   \nonumber \\
   &
  +\eta'-\eta+H'-H
   \hspace{-41mm}
   &
   (\mbox{HFISW+PF})
   \label{E:<Sdelta_l>TinDMB&BKYBwithPF}
\end{align}
with the bond-dependent sums and their averages over bonds
\begin{align}
   &
   \eta'
  \equiv
   \frac{1}{N}\sum_{\nu=1}^N
   \frac{\gamma_{\bm{k}_\nu}e^{i\bm{k}_\nu\cdot\bm{\delta}_l}}
        {2\omega_{\bm{k}_\nu}},
   \\
   &
   \eta
  \equiv
   \frac{\sum_{l=1}^z
         J_{\bm{\delta}_l}\langle\!\langle\mathcal{S}\rangle\!\rangle
         \eta'}
        {\sum_{l=1}^z
         J_{\bm{\delta}_l}\langle\!\langle\mathcal{S}\rangle\!\rangle}
  =\frac{1}{N}\sum_{\nu=1}^N\frac{\gamma_{\bm{k}_\nu}^2}{2\omega_{\bm{k}_\nu}}
  =\epsilon+p\tau,
   \\
   &
   H'
  \equiv
   \frac{1}{N}\sum_{\nu=1}^N
   \frac{\gamma_{\bm{k}_\nu}e^{i\bm{k}_\nu\cdot\bm{\delta}_l}}
        {\omega_{\bm{k}_\nu}}\bar{n}_{\bm{k}_\nu},
   \\
   &
   H
  \equiv
   \frac{\sum_{l=1}^z
         J_{\bm{\delta}_l}\langle\!\langle\mathcal{S}\rangle\!\rangle
         H'}
        {\sum_{l=1}^z
         J_{\bm{\delta}_l}\langle\!\langle\mathcal{S}\rangle\!\rangle}
  =\frac{1}{N}\sum_{\nu=1}^N
   \frac{\gamma_{\bm{k}_\nu}^2}{\omega_{\bm{k}_\nu}}\bar{n}_{\bm{k}_\nu}
  =-I_1+pI_2.
\end{align}
When we set $J$ and $J_\perp$ equal,
Eqs. (\ref{E:<Sdelta_l>'0inDMB&BKYBwithPF})--(\ref{E:<Sdelta_l>TinDMB&BKYBwithPF})
reduce to Eqs. (\ref{E:<S>'0inDMB})--(\ref{E:<S>TinDMB}) and
Eqs. (\ref{E:<S>'0inBKYBwithPF})--(\ref{E:<S>TinBKYBwithPF})
with $\eta'$ and $H'$ coinciding with $\eta$ and $H$, respectively.
\begin{figure}[t]
\centering
\includegraphics[width=85mm]{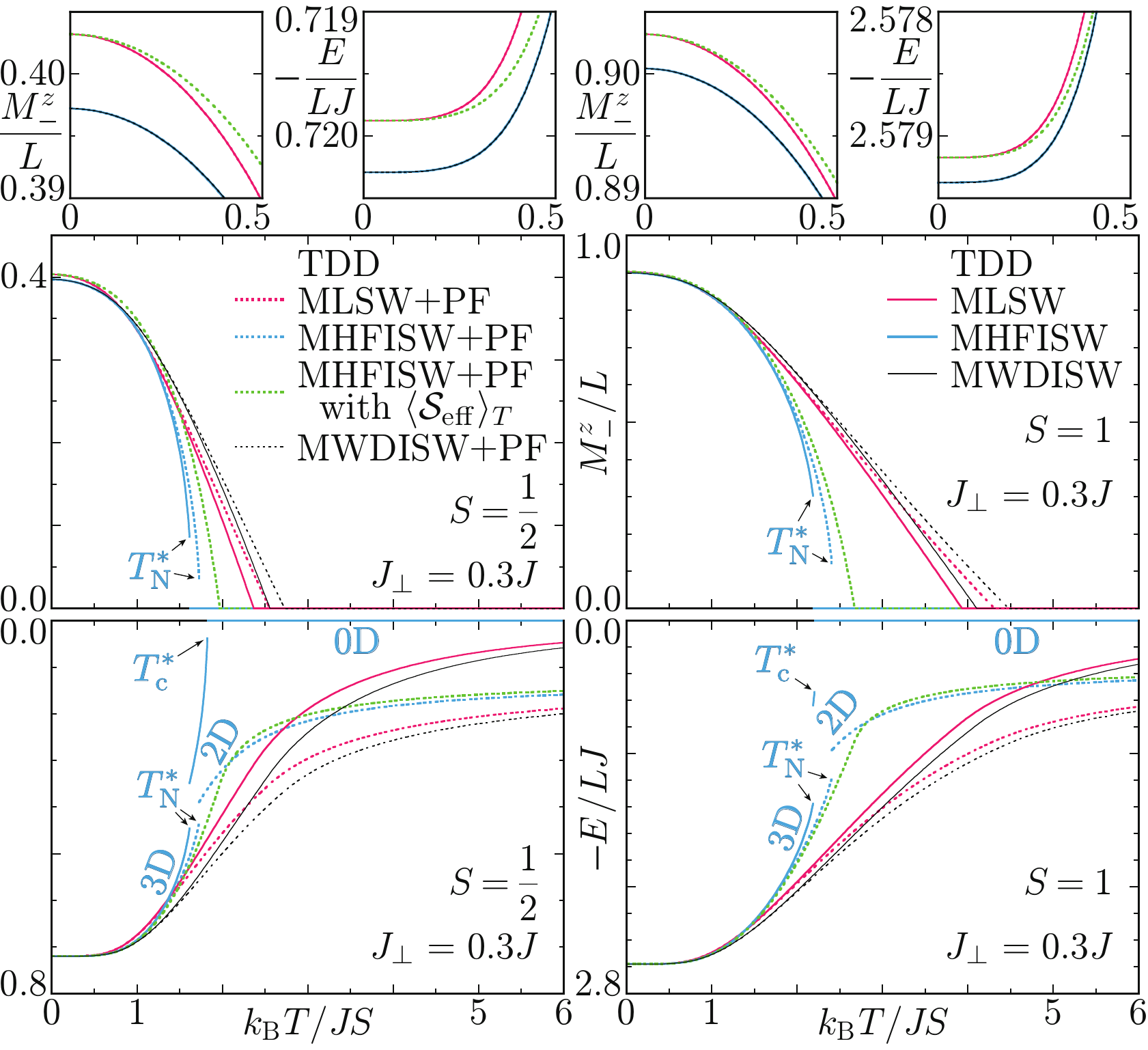}
\vspace*{-2mm}
\caption{(Color online)
         TDD-MSW and TDD-MSW+PF calculations of the staggered magnetization $M_-^z$ and internal
         energy $E$ as functions of temperature for the Hamiltonian (\ref{E:H+Hperp}) of
         $L\rightarrow\infty$ with $J_\perp=0.3J$ in the cases of $S=\frac{1}{2}$ and $S=1$.}
\label{F:MSW&MSW+PFofE&Mst(S=1/2&S=1)3D}
\end{figure}
\begin{table}[t]
\caption{Pseudotransition temperatures, $T_{\mathrm{N}}^*$ and/or $T_{\mathrm{c}}^*$,
         in the TDD-MHFISW and TDD-MHFISW+PF theories
         for the Hamiltonian (\ref{E:H+Hperp}) of $L\rightarrow\infty$ with $J_\perp=0.3J$
         in the cases of $S=\frac{1}{2}$ and $S=1$.
         With temperature reaching $T_{\mathrm{N}}^*$,
         the interlayer coupling $\langle\mathcal{S}_{\bm{\delta}_l=\bm{z}}\rangle_T$ jumps down to
         zero and so does the spontaneous magnetization $M_-^z$,
         and then at the temperature $T_{\mathrm{c}}^*$,
         the intralayer couplings
         $\langle\mathcal{S}_{\bm{\delta}_l=\bm{x}}\rangle_T
         =\langle\mathcal{S}_{\bm{\delta}_l=\bm{y}}\rangle_T$ also jump down to zero to completely
         lose the correlation energy $E$.}
\begin{tabular}{lccccc}
\hline \hline
                & \multicolumn{2}{c}{TDD MHFISW}    &
                & \multicolumn{2}{c}{TDD MHFISW+PF} \\
\cline{2-3} \cline{5-6}
                & $k_\mathrm{B}T_\mathrm{N}^*/JS$ & $k_\mathrm{B}T_\mathrm{c}^*/JS$ &
                & $k_\mathrm{B}T_\mathrm{N}^*/JS$ & $k_\mathrm{B}T_\mathrm{c}^*/JS$ \\
\hline
$S=\frac{1}{2}$ & $1.614$ & $1.822$ &         & $1.726$ &   ---   \\
$S=1$           & $2.187$ & $2.202$ &         & $2.406$ &   ---   \\
\hline \hline
\end{tabular}
\label{T:T_N&T_c}
\end{table}

   We compare the TDD-MSW and TDD-MSW+PF calculations of the staggered magnetization
$M_-^z\equiv\langle\mathcal{M}_-^z\rangle_T$ and internal energy
$E\equiv\langle\mathcal{H}\rangle_T$ in Fig. \ref{F:MSW&MSW+PFofE&Mst(S=1/2&S=1)3D}.
Similarly to the case with single-layer antiferromagnets
(Fig. \ref{F:MSW&MSW+PF&QMCofE&C&chiT(S=1/2&S=1)}), the TDD-MHFISW calculations of
layered antiferromagnets are accompanied by an artificial phase transition of the first order to
the trivial paramagnetic solution.
Every TDD-MHFISW formulation inevitably suffers a pseudotransition from correlated to uncorrelated
spins at a finite temperature, which we shall denote by $T_{\mathrm{c}}^*$ distinguishably from
the real scenario $T_{\mathrm{c}}=\infty$.
However, the present findings require more careful observation and need to be distinguished from
what we have observed in Fig. \ref{F:MSW&MSW+PF&QMCofE&C&chiT(S=1/2&S=1)}.
There are two artificial phase transitions of the first order in the TDD-MHFISW thermodynamics
of layered antiferromagnets (cf. Table \ref{T:T_N&T_c}), unless $J=J_\perp$.
Preceding the total breakdown of the system into free spins at $T=T_{\mathrm{c}}^*$, there occurs
another fictitious phase transition of the first order at a temperature below $T_{\mathrm{c}}^*$
which we shall denote by $T_{\mathrm{N}}^*$, where the spontaneous magnetization
$\langle\mathcal{M}_-^z\rangle_T$ disappears, while the exchange interaction
$\langle\mathcal{H}\rangle_T$ stays finite, losing the interlayer coupling
$\langle\mathcal{S}_{\bm{\delta}_l=\bm{z}}\rangle_T$ but maintaining the intralayer correlation
$\langle\mathcal{S}_{\bm{\delta}_l=\bm{x}}\rangle_T
=\langle\mathcal{S}_{\bm{\delta}_l=\bm{y}}\rangle_T$.
First arises a three-to-two-dimensional phase transition at $T=T_{\mathrm{N}}^*$ and then
an antiferromagnetic-to-paramagnetic phase transition at $T=T_{\mathrm{c}}^*$ on TDD MHFISWs.
That is why we find exactly the same discontinuous transition to occur at exactly the same
temperature in Figs. \ref{F:MSW&MSW+PF&QMCofE&C&chiT(S=1/2&S=1)} and
\ref{F:MSW&MSW+PFofE&Mst(S=1/2&S=1)3D}.
A layered antiferromagnet, once its interlayer coupling is lost, degenerates into planar
antiferromagnets merely stacked up.

   While TDD MHFISWs get free from a total breakdown with the help of auxiliary pseudofermions,
yet they suffer a partial breakdown (cf. Table \ref{T:T_N&T_c}).
Even in the TDD-MHFISW+PF scheme, a layered antiferromagnet is still misled to discontinuously
decouple into planar antiferromagnets at $T=T_{\mathrm{N}}^*$ with its spontaneous magnetization
$\langle\mathcal{M}_-^z\rangle_T$ jumping down to zero.
In order to overcome this difficulty, Irkhin, Katanin, and Katsnelson \cite{I1082} considered
replacing the bond-dependent short-range order parameter (\ref{E:<Sdelta_l>TinDMB&BKYBwithPF}) by
the averaged one
\begin{align}
   \langle\mathcal{S}_{\mathrm{eff}}\rangle_T
   &
  \equiv
   \frac{\sum_{l=1}^z J_{\bm{\delta}_l}\langle\mathcal{S}\rangle_T}
        {\sum_{l=1}^z J_{\bm{\delta}_l}}
   \nonumber \\
   &
  =\frac{J      \langle\mathcal{S}_{\bm{\delta}_l=\bm{x}}\rangle_T
        +J      \langle\mathcal{S}_{\bm{\delta}_l=\bm{y}}\rangle_T
        +J_\perp\langle\mathcal{S}_{\bm{\delta}_l=\bm{z}}\rangle_T}
        {2J+J_\perp}.
   \label{E:<Seff>TinBKYBwithPF}
\end{align}
The thus-tuned MHFISWs+PFs are free from any fictitious transition but misread the ground-state
properties (cf. insets in Fig. \ref{F:MSW&MSW+PFofE&Mst(S=1/2&S=1)3D}).
With $\langle\mathcal{S}\rangle_T$ averaged over $\bm{\delta}_l$'s,
MHFISWs+PFs have the same Bogoliubov transformation (\ref{E:BT}) as MLSWs at $T=0$ and therefore
give the same ground-state energy and magnetization as MLSWs.
Averaging $\langle\mathcal{S}\rangle_T$ spoils the otherwise precise
low-temperature findings of MHFISWs fully demonstrated in
Figs. \ref{F:MSW&QMCofE&C&chi(S=1/2)} and \ref{F:MSW&QMCofE&C&chi(S=1)}.
In addition, every MSW+PF formulation, whether it employs
$\langle\mathcal{S}\rangle_T$ as they are or averages them,
misreads the high-temperature paramagnetic behavior with
everlasting two-dimensional correlation
$\langle\mathcal{S}_{\bm{\delta}_l=\bm{x}}\rangle_T
=\langle\mathcal{S}_{\bm{\delta}_l=\bm{y}}\rangle_T$.

   Indeed the MHFISW+PF thermodynamics formulated with the effective order parameter
(\ref{E:<Seff>TinBKYBwithPF}) is not so precise as the MWDISW one at temperatures far below and
above the ordering temperature $T_{\mathrm{N}}$, but it is highly successful in the truly critical
temperature region near $T_{\mathrm{N}}$.
It reproduces the emergent magnetization $\langle\mathcal{M}_-^z\rangle_T$ much more reasonably
than the MWDISW thermodynamics and can be further improved by taking account of fluctuation
corrections, within a random-phase approximation for instance. \cite{I1082}
Irkhin and Katanin \cite{I12318,I379} further demonstrated that the $O(N)$ model can more suitably
describe the truly critical temperature region.
While the MWDISW thermodynamics is less precise there, yet there may be a possibility of applying
it to layered ferrimagnets, where an anomalous temperature dependence of the spontaneous
magnetization far below $T_{\mathrm{N}}$ is also an interesting topic and remains to be solved.
\cite{K325219,K216003}
Though we have here focused our attention on planar antiferromagnets aiming to solve a longstanding
problem lying in MSW theories for them, the MWDISW scheme can be applied to many other intriguing
magnetic systems including even ``zero-dimensional" clusters \cite{Y157603,C280} to reveal their
overall thermodynamic properties possibly with precise expressions at low temperatures.

\begin{acknowledgments}

   We are grateful to J. Ohara for useful comments.
This study was supported by the Ministry of Education, Culture, Sports,
Science, and Technology of Japan.
\end{acknowledgments}

\appendix*
\renewcommand{\appendixname}{\uppercase{Appendix}}
\section{\uppercase{High-Temperature Asymptotics}}\label{A:HTA}

   We investigate high-temperature asymptotic thermodynamics of square-lattice Heisenberg
antiferromagnets obtained through various MSW formulations.
We employ the effective temperature
$t\equiv{k_{\mathrm{B}}T}/zJ\langle\!\langle\mathcal{S}\rangle\!\rangle$ (\ref{E:t&v}).

\begin{figure}[t]
\centering
\includegraphics[width=85mm]{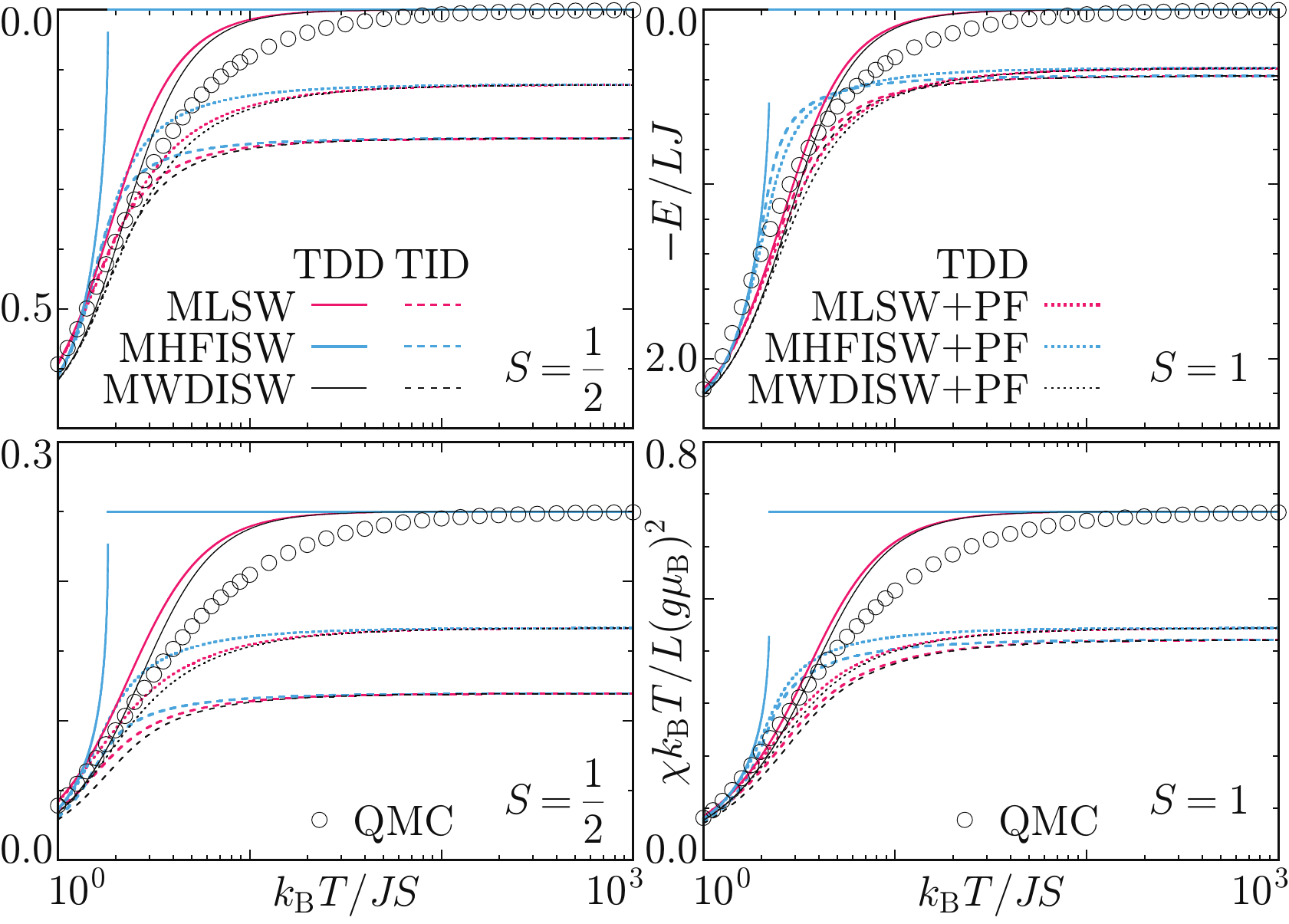}
\vspace*{-2mm}
\caption{(Color online)
         TDD-MSW, TID-MSW, and TDD-MSW+PF calculations of the internal energy $E$ and uniform
         susceptibility $\chi$ as functions of temperature for the Hamiltonian (\ref{E:H}) of
         $L\rightarrow\infty$ in comparison with QMC calculations in the cases of
         $S=\frac{1}{2}$ and $S=1$.}
\label{F:MSW&MSW+PF&QMCofE&chiT(S=1/2&S=1)logT}
\end{figure}

   First, we consider TDD MSWs.
Having in mind that $-1\leq\gamma_{\bm{k}_\nu}\leq 1$, we find that the thermal distribution
function (\ref{E:nkTDD}) no longer depends on the wavevector in the high-temperature limit,
\begin{align}
   \lim_{t\rightarrow\infty}
   \bar{n}_{\bm{k}_\nu}
  =\lim_{t\rightarrow\infty}
   \frac{1}{e^{\omega_{\bm{k}_\nu}/t}-1}
  =\lim_{t\rightarrow\infty}
   \frac{1}{e^{p/t}-1}.
   \label{E:nkTDDMSW(t->oo)}
\end{align}
With converging $\bar{n}_{\bm{k}_\nu}$, i.e. diverging $p$, in the $t\rightarrow\infty$ limit,
we have
\begin{align}
   &
   \lim_{t\rightarrow\infty}\epsilon
  =\frac{1}{2N}\sum_{\nu=1}^N
   \lim_{t\rightarrow\infty}
   \frac{\gamma_{\bm{k}_\nu}^2}{p+\omega_{\bm{k}_\nu}}
  =0,
   \label{E:epsilonTDDMSW(t->oo)}
   \\
   &
   \lim_{t\rightarrow\infty}\tau
  =\frac{1}{2N}\sum_{\nu=1}^N
   \lim_{t\rightarrow\infty}
   \left(
    \frac{p}{\omega_{\bm{k}_\nu}}-1
   \right)
  =0,
   \label{E:tauTDDMSW(t->oo)}
   \\
   &
   \lim_{t\rightarrow\infty}\eta
  =\frac{1}{2N}\sum_{\nu=1}^N
   \lim_{t\rightarrow\infty}
   \frac{\gamma_{\bm{k}_\nu}^2}{\omega_{\bm{k}_\nu}}
  =0,
   \label{E:etaTDDMSW(t->oo)}
   \\
   &
   \lim_{t\rightarrow\infty}H
  =\frac{1}{N}\sum_{\nu=1}^N
   \lim_{t\rightarrow\infty}
   \frac{\gamma_{\bm{k}_\nu}^2}{\omega_{\bm{k}_\nu}}\bar{n}_{\bm{k}_\nu}
  =0,
   \label{E:HTDDMSW(t->oo)}
   \\
   &
   \lim_{t\rightarrow\infty}I_1
  =\frac{1}{N}\sum_{\nu=1}^N
   \lim_{t\rightarrow\infty}
   \omega_{\bm{k}_\nu}\bar{n}_{\bm{k}_\nu}
  =\infty,
   \label{E:I1TDDMSW(t->oo)}
   \\
   &
   \lim_{t\rightarrow\infty}I_2
  =\frac{1}{N}\sum_{\nu=1}^N
   \lim_{t\rightarrow\infty}
   \frac{p}{\omega_{\bm{k}_\nu}}\bar{n}_{\bm{k}_\nu}
  =\lim_{t\rightarrow\infty}
   \bar{n}_{\bm{k}_\nu},
   \label{E:I2TDDMSW(t->oo)}
   \\
   &
   \lim_{t\rightarrow\infty}I_3
  =\frac{1}{N}\sum_{\nu=1}^N
   \lim_{t\rightarrow\infty}
   \bar{n}_{\bm{k}_\nu}(\bar{n}_{\bm{k}_\nu}+1)
   \nonumber \\
   &\qquad\quad
  =\lim_{t\rightarrow\infty}\bar{n}_{\bm{k}_\nu}
   \left(
    \lim_{t\rightarrow\infty}\bar{n}_{\bm{k}_\nu}+1
   \right).
   \label{E:I3TDDMSW(t->oo)}
\end{align}
Then the constraint condition (\ref{E:constraintMst}) becomes
\begin{align}
   S
  =\lim_{t\rightarrow\infty}(\tau+I_2)
  =\lim_{t\rightarrow\infty}\frac{1}{e^{p/t}-1}
   \label{E:constraintMstTDDMSW(t->oo)}
\end{align}
to consistently yield diverging $p$ in the $t\rightarrow\infty$ limit,
\begin{align}
   \lim_{t\rightarrow\infty}\frac{p}{t}
  =\mathrm{ln}\left(1+\frac{1}{S}\right).
   \label{E:p_over_tTDDMSW(t->oo)}
\end{align}
Equations (\ref{E:<S>TinDMB}), (\ref{E:<H(l)>T}), and (\ref{E:chi}) read
\begin{align}
   &
   \lim_{t\rightarrow\infty}
   \frac{E}{NzJ}
 =-\lim_{t\rightarrow\infty}
   (S-\tau-I_2+\eta+H)^2
  =0,
   \label{E:ETDDMSW(t->oo)}
   \\
   &
   \lim_{t\rightarrow\infty}
   \frac{\chi k_{\mathrm{B}}T}{L(g\mu_{\mathrm{B}})^2}
  =\frac{1}{3}\lim_{t\rightarrow\infty}
   \left[
    2(S-\tau-I_2)\vphantom{\frac{p-1}{q}}
   \right.
   \nonumber \\
   &\qquad\times
   \left.
    \frac{p-1}{q}\left(\bar{n}_{\bm{0}}+\frac{1}{2}\right)
   +I_3
   \right]
  =\frac{S(S+1)}{3}.
   \label{E:chiTTDDMSW(t->oo)}
\end{align}
We learn that TDD MSWs hit the correct high-temperature limit.
TDD MHFISWs end up with an artificial transition to the paramagnetic solution,
but TDD MLSWs and TDD MWDISWs really give correct high-temperature asymptotics,
as is demonstrated in Fig. \ref{F:MSW&MSW+PF&QMCofE&chiT(S=1/2&S=1)logT}.

   Next, we consider TID MSWs.
Having in mind that $p=1$ as well as $-1\leq\gamma_{\bm{k}_\nu}\leq 1$, we find that the thermal
distribution function (\ref{E:nkTID}) still depends on the wavevector in the high-temperature
limit,
\begin{align}
   \lim_{t\rightarrow\infty}
   \bar{n}_{\bm{k}_\nu}
   &
  =\lim_{t\rightarrow\infty}
   \frac{1}
        {e^{\omega_{\bm{k}_\nu}/t
           -\mu/tzJ\langle\!\langle\mathcal{S}\rangle\!\rangle\omega_{\bm{k}_\nu}}-1}
   \nonumber \\
   &
  =\lim_{t\rightarrow\infty}
   \frac{1}
        {e^{-\mu/tzJ\langle\!\langle\mathcal{S}\rangle\!\rangle\omega_{\bm{k}_\nu}}-1},
   \label{E:nkTIDMSW(t->oo)}
\end{align}
making it difficult to analytically calculate the high-temperature asymptotics.
$\epsilon$, $\tau$, and $\eta$ are not dependent on temperature.
With converging $\bar{n}_{\bm{k}_\nu}$ in the $t\rightarrow\infty$ limit, we have
\begin{align}
   &
   \lim_{t\rightarrow\infty}H
  =\frac{1}{N}\sum_{\nu=1}^N
   \frac{\gamma_{\bm{k}_\nu}^2}{\omega_{\bm{k}_\nu}}
   \lim_{t\rightarrow\infty}
   \bar{n}_{\bm{k}_\nu},
   \label{E:HTIDMSW(t->oo)}
   \\
   &
   \lim_{t\rightarrow\infty}I_1
  =\frac{1}{N}\sum_{\nu=1}^N
   \omega_{\bm{k}_\nu}
   \lim_{t\rightarrow\infty}
   \bar{n}_{\bm{k}_\nu},
   \label{E:I1TIDMSW(t->oo)}
   \\
   &
   \lim_{t\rightarrow\infty}I_2
  =\frac{1}{N}\sum_{\nu=1}^N
   \frac{1}{\omega_{\bm{k}_\nu}}
   \lim_{t\rightarrow\infty}
   \bar{n}_{\bm{k}_\nu},
   \label{E:I2TIDMSW(t->oo)}
   \\
   &
   \lim_{t\rightarrow\infty}I_3
  =\frac{1}{N}\sum_{\nu=1}^N
   \lim_{t\rightarrow\infty}
   \bar{n}_{\bm{k}_\nu}(\bar{n}_{\bm{k}_\nu}+1).
   \label{E:I3TIDMSW(t->oo)}
\end{align}
We numerically solve the constraint condition (\ref{E:constraintMst}) in the high-temperature limit
\begin{align}
   S
   &
  =\lim_{t\rightarrow\infty}(\tau+I_2)
   \nonumber \\
   &
  =\tau_{p=1}
  +\frac{1}{N}\sum_{\nu=1}^N
   \frac{1}{\omega_{\bm{k}_\nu}}
   \lim_{t\rightarrow\infty}
   \frac{1}
        {e^{-\mu/tzJ\langle\!\langle\mathcal{S}\rangle\!\rangle\omega_{\bm{k}_\nu}}-1}
   \label{E:constraintMstTIDMSW(t->oo)}
\end{align}
to find $\lim_{t\rightarrow\infty}(-\mu/tzJ\langle\!\langle\mathcal{S}\rangle\!\rangle)$ to be
nonzero and thus $\lim_{t\rightarrow\infty}\bar{n}_{\bm{k}_\nu}$ to be finite.
Equations (\ref{E:<S>TinDMB}), (\ref{E:<H(l)>T}), and (\ref{E:chi}) read
\begin{align}
   &
   \lim_{t\rightarrow\infty}
   \frac{E}{NzJ}
 =-\left(
    \eta_{p=1}
   +\frac{1}{N}\sum_{\nu=1}^N
    \frac{\gamma_{\bm{k}_\nu}^2}{\omega_{\bm{k}_\nu}}
   \right.
   \nonumber \\
   &\qquad\qquad\qquad\times
   \left.
    \vphantom{\sum_{\nu=1}^N}
    \lim_{t\rightarrow\infty}
    \frac{1}
         {e^{-\mu/tzJ\langle\!\langle\mathcal{S}\rangle\!\rangle\omega_{\bm{k}_\nu}}-1}
   \right)^2,
   \label{E:ETIDMSW(t->oo)}
   \\
   &
   \lim_{t\rightarrow\infty}
   \frac{\chi k_{\mathrm{B}}T}{L(g\mu_{\mathrm{B}})^2}
  =\frac{1}{3N}\sum_{\nu=1}^N
   \lim_{t\rightarrow\infty}
   \frac{1}
        {e^{-\mu/tzJ\langle\!\langle\mathcal{S}\rangle\!\rangle\omega_{\bm{k}_\nu}}-1}
   \nonumber \\
   &\qquad\qquad\qquad\times
   \left(
    \frac{1}
         {e^{-\mu/tzJ\langle\!\langle\mathcal{S}\rangle\!\rangle\omega_{\bm{k}_\nu}}-1}
   +1
   \right).
   \label{E:chiTTIDMSW(t->oo)}
\end{align}
We find
$\lim_{t\rightarrow\infty}(-\mu/tzJ\langle\!\langle\mathcal{S}\rangle\!\rangle)$
to be $1.31972709$ and $0.74682827$,
$\lim_{t\rightarrow\infty}E/LJ$
to be $-0.21475012$ and $-0.37814423$, and
$\lim_{t\rightarrow\infty}\chi k_{\mathrm{B}}T/L(g\mu_{\mathrm{B}})^2$
to be $0.11970015$ and $0.42202662$ for $S=\frac{1}{2}$ and $S=1$, respectively
(See Fig. \ref{F:MSW&MSW+PF&QMCofE&chiT(S=1/2&S=1)logT}).
We learn that TID MSWs cannot give correct high-temperature asymptotics.

   Finally, we consider TDD MSWs combined with PFs.
The Bar'yakhtar-Krivoruchko-Yablonski{\u\i} thermal average of MSWs (\ref{E:nkTDDinBKYBwithPF})
in the high-temperature limit looks the same as (\ref{E:nkTDDMSW(t->oo)}).
That of PFs (\ref{E:fTDDinBKYBwithPF}) is free from wavevector dependence at every temperature,
\begin{align}
   \bar{f}
  =\frac{1}{1-e^{\delta\varepsilon/tzJ\langle\!\langle\mathcal{S}\rangle\!\rangle}}
  =\frac{1}{1-e^{(2S+1)p/t}}.
   \label{E:fTDDMSW+PF(t->oo)}
\end{align}
If the calculations of the high-temperature limit
(\ref{E:epsilonTDDMSW(t->oo)})--(\ref{E:I3TDDMSW(t->oo)}) remain unchanged,
the constraint condition (\ref{E:constraintMstinBKYBwithPF}) becomes
\begin{align}
   S
   &
  =\lim_{t\rightarrow\infty}
   [\tau+I_2+(2S+1)\bar{f}]
   \nonumber \\
   &
  =\lim_{t\rightarrow\infty}
   \left[
    \frac{1}{e^{p/t}-1}-\frac{2S+1}{e^{(2S+1)p/t}-1}
   \right]
   \nonumber \\
   &
  \equiv
   S-SB_S\left(S\lim_{t\rightarrow\infty}\frac{p}{t}\right)
   \nonumber \\
   &
  =S
  -\frac{S(S+1)}{3}
   \lim_{t\rightarrow\infty}\frac{p}{t}
  +O\left[\left(\lim_{t\rightarrow\infty}\frac{p}{t}\right)^3\right]
   \label{E:constraintMstTDDMSW+PF(t->oo)X}
\end{align}
with the Brillouin function
\begin{align}
   B_S(x)
  \equiv
   \frac{2S+1}{2S}
   \mathrm{coth}\frac{2S+1}{2S}x
  -\frac{1}{2S}
   \mathrm{coth}\frac{1}{2S}x.
   \label{E:BF}
\end{align}
Since $B_S(x)$ is a monotonically increasing function, $\lim_{t\rightarrow\infty}p/t=0$ is one
and only real solution of Eq. (\ref{E:constraintMstTDDMSW+PF(t->oo)X}), and therefore,
the Bar'yakhtar-Krivoruchko-Yablonski{\u\i} thermal averages (\ref{E:nkTDDMSW(t->oo)}) and
(\ref{E:fTDDMSW+PF(t->oo)}) both diverge in the $t\rightarrow\infty$ limit.
Then, the calculations (\ref{E:HTDDMSW(t->oo)})--(\ref{E:I3TDDMSW(t->oo)}) are no longer reliable,
because every thermal-distribution-weighted integration over $\bm{k}$ is not necessarily
exchangeable with the $t\rightarrow\infty$ operation.
Yet applying Eqs. (\ref{E:HTDDMSW(t->oo)})--(\ref{E:I3TDDMSW(t->oo)}) to
Eqs. (\ref{E:<S>TinBKYBwithPF}), (\ref{E:EinBKYBwithPF}), and (\ref{E:chiinBKYBwithPF}) yields
\begin{align}
   &
   \lim_{t\rightarrow\infty}
   \frac{E}{NzJ}
 =-\lim_{t\rightarrow\infty}
   [S-\tau-I_2-(2S+1)\bar{f}
   \nonumber \\
   &\qquad
   +\eta+H]^2
  =0,
   \label{E:ETDDMSW+PF(t->oo)}
   \\
   &
   \lim_{t\rightarrow\infty}
   \frac{\chi k_{\mathrm{B}}T}{L(g\mu_{\mathrm{B}})^2}
  =\frac{1}{3}\lim_{t\rightarrow\infty}
   \left\{
    2[S-\tau-I_2-(2S+1)\bar{f}]
    \vphantom{\frac{p-1}{q}}
   \right.
   \nonumber \\
   &\qquad\times
   \left.
    \frac{p-1}{q}
    \left(\bar{n}_{\bm{0}}+\frac{1}{2}\right)
   +I_3+(2S+1)^2\bar{f}(1-\bar{f})
   \right\}
   \nonumber \\
   &\qquad
  =\frac{S^2}{3}B_S'\left(S\lim_{t\rightarrow\infty}\frac{p}{t}\right)
  =\frac{S(S+1)}{9},
   \label{E:chiTTDDMSW+PF(t->oo)}
\end{align}
which are indeed inconsistent with the correct calculations demonstrated in
Fig. \ref{F:MSW&MSW+PF&QMCofE&chiT(S=1/2&S=1)logT}.

   In order to analyze the high-temperature asymptotics of the TDD-MSW+PF thermodynamics correctly,
we expand every thermal quantity into high-temperature series.
Employing the state-density function (\ref{E:w(x)}) and its explicit expression
(\ref{E:w(x)(L->oo)}) again, the thermal-distribution-weighted sums read
\begin{align}
   I_1
   &
  =\int_0^{p-q}
   \frac{x+q}{e^{(x+q)/t}-1}w(x)dx
  =p\left(\frac{2}{\pi}\right)^2
   \nonumber \\
   &\times
   \int_0^1
   \frac{\sqrt{1-(y/p)^2}K\bigl(\sqrt{1-y^2}\bigr)}
        {e^{(p/t)\sqrt{1-(y/p)^2}}-1}
   dy,
   \label{E:I1int}
   \\
   I_2
   &
  =\int_{0}^{p-q}
   \frac{p/(x+q)}{e^{(x+q)/t}-1}
   w(x)dx
  =\left(\frac{2}{\pi}\right)^2
   \nonumber \\
   &\times
   \int_0^1
   \frac{K\bigl(\sqrt{1-y^2}\bigr)/\sqrt{1-(y/p)^2}}
        {e^{(p/t)\sqrt{1-(y/p)^2}}-1}
   dy,
   \label{E:I2int}
   \\
   I_3
   &
  =\int_{0}^{p-q}
   \frac{e^{(x+q)/t}}
        {\bigl[e^{(x+q)/t}-1\bigr]^{2}}
   w(x)dx
  =\left(\frac{2}{\pi}\right)^2
   \nonumber \\
   &\times
   \int_0^1
   \frac{e^{(p/t)\sqrt{1-(y/p)^2}}K\bigl(\sqrt{1-y^2}\bigr)}
        {\bigl[e^{(p/t)\sqrt{1-(y/p)^2}}-1\bigr]^2}
   dy.
   \label{E:I3int}
\end{align}
At low temperatures, having in mind that low-energy excitations, i.e. $x$ much smaller than $p-q$,
make major contributions to the integrals (\ref{E:I1int})--(\ref{E:I3int}), we expand $w(x)$ in
powers of $x$ as Eqs. (\ref{E:w(x)(L->oo)}) and (\ref{E:w_l}), while at high temperatures of
current interest, having in mind that
\begin{align}
   &
   \left(
    \frac{2}{\pi}
   \right)^2
   \int_0^1
   y^{2l} K\bigl(\sqrt{1-y^2}\bigr) dy
  =\frac{1}{\pi}\left[\frac{\varGamma(l+1/2)}{\varGamma(l+1)}\right]^2
   \nonumber \\
   &\qquad
  =\left[\frac{(2l-1)!!}{2^l l!}\right]^2\ 
   (l=0,1,2,\cdots),
   \label{E:Kint=>Gamma}
\end{align}
we expand thermal distribution functions as
\begin{align}
   &
   \frac{1}{e^\beta-1}
  =\frac{1}{\beta}-\frac{1}{2}+\frac{\beta}{12}-\frac{\beta^3}{720}+O(\beta^5),
   \nonumber \\
   &
   \frac{e^\beta}{(e^\beta-1)^2}
 =-\frac{\partial}{\partial\beta}\frac{1}{e^\beta-1}.
   \label{E:n&n(n+1)&f&f(1-f)expansions}
\end{align}
Note further that
\begin{align}
   p-2\epsilon
   &
  =\int_0^{p-q}(x+q)w(x)dx
  =p\left(\frac{2}{\pi}\right)^2
   \nonumber \\
   &\times
   \int_0^1
   \sqrt{1-(y/p)^2}
   K\bigl(\sqrt{1-y^2}\bigr)dy,
   \label{E:pint}
   \\
   2\tau+1
   &
  =\int_0^{p-q}\frac{p}{x+q}w(x)dx
  =\left(\frac{2}{\pi}\right)^2
   \nonumber \\
   &\times
   \int_0^1
   \frac{K\bigl(\sqrt{1-y^2}\bigr)}{\sqrt{1-(y/p)^2}}dy.
   \label{E:tauint}
\end{align}
Supposing $p$ to be of order $\sqrt{t}$, we consider such high temperatures as to satisfy
$1\ll p\ll t$.
Via the expansions (\ref{E:n&n(n+1)&f&f(1-f)expansions}) with
$\beta\equiv(p/t)\sqrt{1-(y/p)^2}$, the integrals (\ref{E:I1int})--(\ref{E:I3int}) become
\begin{align}
   I_1
   &
  =p\left(\frac{2}{\pi}\right)^2
   \int_0^1
   \left[
    \frac{t}{p}+\frac{p}{12t}+O(t^{-3/2})
   \right]
   K\bigl(\sqrt{1-y^2}\bigr)dy
   \nonumber \\
   &\ 
  -\frac{p-2\epsilon}{2}
  =t
  -\frac{p}{2}
  +\frac{p^2}{12t}
  +\epsilon
  +O(t^{-1}),
   \label{E:I1int(TDDMSW+PF)HTSE}
   \\
   I_2
   &
  =\left(\frac{2}{\pi}\right)^2
   \int_0^1
   \left\{
    \frac{t}{p}
    \left[
     1+\left(\frac{y}{p}\right)^2
    \right]
   +\frac{p}{12t}
   +O(t^{-3/2})
   \right\}
   \nonumber \\
   &\ \times
   K\bigl(\sqrt{1-y^2}\bigr)dy
  -\frac{2\tau+1}{2}
   \nonumber \\
   &
  =\frac{t}{p}
  -\frac{1}{2}
  +\frac{t}{4p^3}+\frac{p}{12t}
  -\tau
  +O(t^{-3/2}),
   \label{E:I2int(TDDMSW+PF)HTSE}
   \\
   I_3
   &
  =\left(\frac{2}{\pi}\right)^2
   \int_0^1
   \left\{
    \frac{t^2}{p^2}
    \left[
     1+\left(\frac{y}{p}\right)^2+\left(\frac{y}{p}\right)^4
    \right]
   \right.
   \nonumber \\
   &
   \left.
    \vphantom{\left(\frac{y}{p}\right)^4}
   -\frac{1}{12}
   +\frac{p^2}{240t^2}
   +O(t^{-2})
   \right\}
   K\bigl(\sqrt{1-y^2}\bigr)dy
   \nonumber \\
   &
  =\frac{t^2}{p^2}
  +\frac{t^2}{4p^4}-\frac{1}{12}
  +\frac{9t^2}{64p^6}+\frac{p^2}{240t^2}
  +O(t^{-2}),
   \label{E:I3int(TDDMSW+PF)HTSE}
\end{align}
where the $p$ dependences of $\epsilon$ and $\tau$ read
\begin{align}
   &\!\!\!
   p-2\epsilon
  =p\left(\frac{2}{\pi}\right)^2
   \int_0^1
   \left[
    1-\frac{y^2}{2p^2}-\frac{y^4}{8p^4}
   +O\left(\frac{y^6}{p^6}\right)
   \right]
   \nonumber \\
   &\!\!\!\,\times
   K\bigl(\sqrt{1-y^2}\bigr)dy
  =p-\frac{1}{8p}-\frac{9}{512p^3}+O(p^{-5}),
   \label{E:epsilon(TDDMSW+PF)LpSE}
   \\
   &\!\!\!
   2\tau+1
  =\left(\frac{2}{\pi}\right)^2
   \int_0^1
   \left[
    1+\frac{y^2}{2p^2}+\frac{3y^4}{8p^4}
   +O\left(\frac{y^6}{p^6}\right)
   \right]
   \nonumber \\
   &\!\!\!\,\times
   K\bigl(\sqrt{1-y^2}\bigr)dy
  =1+\frac{1}{8p^2}+\frac{27}{512p^4}+O(p^{-6}).
   \label{E:tau(TDDMSW+PF)LpSE}
\end{align}
Putting $\beta\equiv(2S+1)p/t$ in Eq. (\ref{E:n&n(n+1)&f&f(1-f)expansions}) yields
the high-temperature series expansions of $-\bar{f}$ and $-\bar{f}(1-\bar{f})$.
The constraint condition (\ref{E:constraintMstinBKYBwithPF}) becomes
\begin{align}
   \frac{S(S+1)p}{3t}-\frac{t}{4p^3}+O(t^{-3/2})
  =0
   \label{E:constraintMstTDDMSW+PF(t->oo)O}
\end{align}
to reveal that $p$ indeed diverges as $\sqrt{t}$ at high temperatures,
\begin{align}
   p
  =\left[
    \frac{3}{4S(S+1)}
   \right]^{1/4}
   \sqrt{t}
  +O(t^{-1/2}).
   \label{E:p(TDDMSW+PF)HTSE}
\end{align}
Equations (\ref{E:<S>TinBKYBwithPF}) and (\ref{E:chiinBKYBwithPF}) read
\begin{align}
   &\!\!\!\!
   \langle\mathcal{S}\rangle_T
  =\frac{1}{2}\sqrt{\frac{S(S+1)}{3}}+O(t^{-1}),
   \label{E:<S>T(TDDMSW+PF)HTSE}
   \\
   &\!\!\!\!
   \chi^{xx}=\chi^{yy}=0,\ 
   \frac{\chi^{zz}k_{\mathrm{B}}T}{L(g\mu_{\mathrm{B}})^2}
  =\frac{2S(S+1)}{3}+O(t^{-1}).
   \label{E:chizzT(TDDMSW+PF)HTSE}
\end{align}
Having in mind that
\begin{align}
   t
  =\frac{k_\mathrm{B}T}{zJS}
   \times
   \left\{
   \!\!
   \begin{array}{lr}
    \displaystyle
    1
    & (\mbox{MLSW+PF}) \\
    \displaystyle
    \left[1+O(T^{-1/2})\right]
    & (\mbox{MWDISW+PF}) \\
    \displaystyle
    \left[
     \sqrt{\frac{12S}{S+1}}
    +O(T^{-1})
    \right]
    & (\mbox{MHFISW+PF}) \\
   \end{array}
   \right.\!\!\!,
   \label{E:t(T)TDDMSW+PF}
\end{align}
we eventually have
\begin{align}
   &
   \frac{E}{NzJ}
 =-\frac{S(S+1)}{12}+O(T^{-1}),
   \label{E:E(TDDMSW+PF)HTSE}
   \\
   &
   \frac{\chi k_{\mathrm{B}}T}{L(g\mu_{\mathrm{B}})^2}
  =\frac{2S(S+1)}{9}+O(T^{-1}),
   \label{E:chiT(TDDMSW+PF)HTSE}
\end{align}
which are now consistent with the numerical calculations demonstrated in
Fig. \ref{F:MSW&MSW+PF&QMCofE&chiT(S=1/2&S=1)logT}.
We learn that TDD MSWs+PFs cannot give correct high-temperature asymptotics.

\end{document}